\documentclass[%
nofootinbib,
amsmath,2amssymb,
prd,
twocolumn,
10pt
]{revtex4-1}
\usepackage[bottom]{footmisc}
\usepackage[utf8]{inputenc}
\usepackage{slashed}
\usepackage[%dvipdfm,
colorlinks=true,
linkcolor=blue,
breaklinks=true,
urlcolor=blue,
citecolor=blue]{hyperref}
\usepackage{multirow}
\usepackage{diagbox}
\usepackage{dcolumn}% Align table columns on decimal point
\usepackage{bm}% bold math
\usepackage{hyperref}% add hypertext capabilities
\usepackage{setspace}
\usepackage{makecell}
\usepackage{threeparttable}
\usepackage{footnote}
\usepackage{graphicx}% Include figure files
\usepackage{float}
\allowdisplaybreaks[4]

\newcommand{\ket}[1]{\left| #1 \right\rangle}

\newcommand{\lra}[1]{\left(#1\right)}

\usepackage{relsize}
\def\babar{\mbox{\slshape B\kern-0.1em{\smaller A}\kern-0.1em
    B\kern-0.1em{\smaller A\kern-0.2em R}}}

\begin{document}

\title{A survey of heavy-heavy hadronic molecules}

\author{Xiang-Kun Dong$^{1,2}$}\email{dongxiangkun@itp.ac.cn}
\author{Feng-Kun Guo$^{1,2}$} \email{fkguo@itp.ac.cn}
\author{Bing-Song Zou$^{1,2,3}$}\email{zoubs@itp.ac.cn}

\address{%
$^1$CAS Key Laboratory of Theoretical Physics, Institute of Theoretical Physics,\\
Chinese Academy of Sciences, Beijing 100190, China\\
$^2$School of Physical Sciences, University of Chinese Academy of Sciences, Beijing 100049, China\\
$^3$School of Physics, Central South University, Changsha 410083, China
}

\begin{abstract}
The spectrum of hadronic molecules composed of heavy-antiheavy charmed hadrons has been obtained in our previous work. The potentials are constants at the leading order, which are estimated from resonance saturation. The experimental candidates of hadronic molecules, say $X(3872)$, $Y(4260)$, three $P_c$ states and $P_{cs}(4459)$, fit the spectrum well. The success in describing the pattern of heavy-antiheavy hadronic molecules stimulates us to give more predictions for the heavy-heavy cases, which are less discussed in literature than the heavy-antiheavy ones. Given that the heavy-antiheavy hadronic molecules, several of which have strong experimental evidence, emerge from the dominant constant interaction from resonance saturation, we find that the existence of many heavy-heavy hadronic molecules is natural. Among these predicted heavy-heavy states we highlight the $DD^*$ molecule and the $D^{(*)}\Sigma_c^{(*)}$ molecules, which are the partners of the famous $X(3872)$ and $P_c$ states. Quite recently, LHCb collaboration reported a doubly charmed tetraquark state, $T_{cc}$, which is in line with our results for the $DD^*$ molecule. 
With the first experimental signal of this new kind of exotic states, the upcoming update of the LHCb experiment as well as other experiments will provide more chances of observing the heavy-heavy hadronic molecules.

\end{abstract}

\maketitle

\medskip

% \tableofcontents

\medskip

\section{Introduction}\label{sec:1}

The fact that quantum  chromodynamics (QCD) is nonperturbative at low energy makes the calculation of the whole hadron spectrum from first-principle too difficult at the present stage. The quark model proposed in Refs.~\cite{GellMann:1964nj,Zweig:1964jf} successfully classify plenty of hadrons and its later developments after the birth of QCD (see, e.g., Refs.~\cite{Godfrey:1985xj,Capstick:1985xss}) provide a remarkable description of the hadron spectrum of $q\bar q$ mesons and $qqq$ baryons. However, there was little clear evidence for the multiquark states predicted in Ref.~\cite{GellMann:1964nj} until the discovery of the $X(3872)$~\cite{Choi:2003ue}, also known as $\chi_{c1}(3872)$ in the Review of Particle Physics (RPP)~\cite{ParticleDataGroup:2020ssz}. Since then many near-threshold structures, e.g. the $Z_c(3900)^\pm$~\cite{Ablikim:2013mio,Liu:2013dau,Ablikim:2013xfr}, the $Z_c(4020)^\pm$~\cite{Ablikim:2013emm,Ablikim:2013wzq}, the $Z_b(10610/10650)^\pm$~\cite{Belle:2011aa,Garmash:2015rfd}, the $Z_{cs}(3985)^-$~\cite{Ablikim:2020hsk}  and the $P_c$ states~\cite{Aaij:2019vzc}, have been observed in the worldwide high energy experiments. These so-called exotic states are clearly outside the scope of the traditional quark model consisting of $q\bar q$ mesons and $qqq$ baryons but their inner structures are still under debate (see Refs.~\cite{Chen:2016qju,Hosaka:2016pey,Richard:2016eis,Lebed:2016hpi,Esposito:2016noz,Guo:2017jvc,Ali:2017jda,Olsen:2017bmm,Kou:2018nap,Kalashnikova:2018vkv,Cerri:2018ypt,Liu:2019zoy,Brambilla:2019esw,Guo:2019twa,Yang:2020atz,Ortega:2020tng,Dong:2021juy} for recent reviews of the multiquark states).

One peculiar and thus important property of these exotic states mentioned above is that they are all located quite close to the thresholds of a pair of hadrons that they can couple to.  Therefore it is natural to consider them as hadronic molecules\footnote{Note that the concept of hadronic molecules have been extended from bound states~\cite{Weinberg:1962hj,Weinberg:1963zza,Weinberg:1965zz} to near-threshold resonances~\cite{Baru:2003qq} and virtual states~\cite{Matuschek:2020gqe}. } composed of the corresponding hadron pairs and such an idea was widely explored in the literature, see Ref.~\cite{Guo:2017jvc} for a review of hadronic molecules and Ref.~\cite{Dong:2020hxe} for a discussion of the general behaviors of near-threshold structures. Although many works have used different methods to explain certain properties of these near-threshold states, a whole and systematic spectrum of hadronic molecules will deepen our understanding of these exotic states. In Ref.~\cite{Dong:2021juy}, we have provided such a spectrum for heavy-antiheavy (taking charmed hadrons for example) hadronic molecules by solving the single channel Bethe-Salpeter (BS) equation with constant interactions, which are assumed  to  be vector meson-exchange  saturated. In this work we extend the work in Ref.~\cite{Dong:2021juy} to the heavy-heavy systems to make the spectrum more complete.

Note that all the experimentally established exotic states mentioned above are hidden-charm or hidden-bottom ones. It is much more difficult to produce the states with doubly heavy quarks than those with heavy-antiheavy quarks since that require at least two heavy quark-antiquark pairs to be produced and two heavy (anti)quarks to move in a particular phase space region. It is only until very recently that the first experimental evidence for a double-charm tetraquark state was reported~\cite{LHCb:2021vvq,LHCb:2021auc}, which immediately stimulated a series of theoretical studies~\cite{Li:2021zbw,Agaev:2021vur,Ling:2021bir,Meng:2021jnw,Chen:2021vhg,Feijoo:2021ppq,Yan:2021wdl,Wang:2021yld,Xin:2021wcr,Fleming:2021wmk,Azizi:2021aib,Chen:2021tnn,Ren:2021dsi,Jin:2021cxj}. Despite that, many attempts have been made to investigate the possible states with doubly heavy quarks, see the discussions in Sec.~\ref{sec:4}. On the one hand, the heavy antiquark–diquark symmetry for antiheavy baryons may lead to the corresponding doubly-heavy tetraquark states. On the other hand, like the heavy-antiheavy systems, the heavy-heavy hadronic molecules are also expected to exist.

This work is organized as follows. In Sec.~\ref{sec:2}, we give a brief summary of the interactions between heavy hadrons following the heavy quark spin symmetry. In Sec.~\ref{sec:3}, the potentials resulting from these interactions are presented and the molecular states obtained by solving the single channel BS equation are listed. Some selected states are discussed in Sec.~\ref{sec:4}, together with an incomplete review on the status of such systems in the literature. We then close this paper with a brief summary in Sec.~\ref{sec:5}.

\section{Lagrangian from heavy quark spin symmetry}\label{sec:2}

The interactions between hadrons consisting of one or more heavy quarks, say $c$ and $b$ due to their much larger masses than the typical QCD scale, can be constructed systematically under the guidance of heavy quark symmetries\footnote{For the spirits of heavy quark symmetries, we refer to the pioneer works~\cite{Shifman:1986sm,Politzer:1988bs,Politzer:1988wp,Isgur:1989ed,Isgur:1989vq} and reviews~\cite{Neubert:1993mb,Manohar:2000dt}} and such Lagrangians can be found in, e.g. Refs.~\cite{Wise:1992hn,Casalbuoni:1992gi,Casalbuoni:1996pg,Grinstein:1992qt,Falk:1991nq,Falk:1992cx,Yan:1992gz,Casalbuoni:1992dx,Casalbuoni:1996pg,Liu:2011xc}. The relevant Lagrangians and some details have been collected in our previous work~\cite{Dong:2021juy} and here we will just give the final results that are needed. Note that in the following we are interested in the potential near threshold and will not consider coupled channels. Therefore, we only consider the coupling between heavy hadrons and light vector mesons and omit the Lagrangian that results in potentials proportional to the square of transferred momentum, $\bm q^2$.

\subsection{Coupling of light vector mesons and heavy mesons}

The coupling of heavy mesons and light vector mesons can be introduced by using the hidden local symmetry approach~\cite{Bando:1984ej,Bando:1987br,Meissner:1987ge} and the leading order Lagrangian is expressed as~\cite{Casalbuoni:1992dx,Casalbuoni:1996pg,Dong:2021juy}
\begin{align}
&\mathcal L_{PPV}=-\sqrt{2}\beta g_V \left(P_a^{(Q)}P_b^{(Q)\dagger}-P_b^{(\bar Q)}P_a^{(\bar Q)\dagger}\right)v_\mu V^\mu_{ab}\notag\\
&+\sqrt{2}\beta g_V \left(P_a^{*(Q)\nu}P_{b\nu}^{*(Q)\dagger}-P_b^{*(\bar Q)\nu}P_{a\nu}^{*(\bar Q)\dagger}\right)v_\mu V^\mu_{ab}\notag\\
&-\sqrt{2}\beta_2 g_V \left(P_{1a}^{(Q)\nu}P_{1b\nu}^{(Q)\dagger}-P_{1b}^{(\bar Q)\nu}P_{1a\nu}^{(\bar Q)\dagger}\right)v_\mu V^\mu_{ab}\notag\\
&+\sqrt{2}\beta_2 g_V \left(P_{2a}^{(Q)\alpha\beta}P_{2b\alpha\beta}^{(Q)\dagger}-P_{2b}^{(\bar Q)\alpha\beta}P_{2a\alpha\beta}^{(\bar Q)\dagger}\right)v_\mu V^\mu_{ab}\notag\\
&+\Big[\sqrt2\zeta_1 g_V \left(P_{2a}^{(Q)\mu\nu}P_{b\nu}^{(Q)*\dagger}+P_{2b}^{(\bar Q)\mu\nu}P_{a\nu}^{(\bar Q)*\dagger}\right)V_{ab\mu}\notag\\
&-\frac{i\zeta_1 g_V}{\sqrt3} \epsilon_{\alpha\beta\gamma\delta}\left(P_{1a}^{(Q)\alpha} P_b^{(Q)*\dagger\beta}+P_{1b}^{(\bar Q)\alpha}  P_a^{(\bar Q)*\dagger\beta}\right)v^\gamma V_{ab}^\delta\notag\\
&-\frac{2\zeta_1 g_V}{\sqrt3} \left(P_{1a\mu}^{(Q)}P_b^{(Q)\dagger}-P_{1b\mu}^{(\bar Q) }P_a^{(\bar Q)\dagger}\right) V_{ab}^\mu+{\rm h.c.}\Big],\label{eq:LPPV}
\end{align}
where $a,b$ are the SU(3) flavor indices and 
\begin{align}
 V=\left(\begin{array}{ccc}
\frac{\omega}{\sqrt{2}}+\frac{\rho^{0}}{\sqrt{2}} & \rho^{+} & K^{*+} \\
\rho^{-} & \frac{\omega}{\sqrt{2}}-\frac{\rho^{0}}{\sqrt{2}} & K^{* 0} \\
K^{*-} & \bar{K}^{* 0} & \phi
\end{array}\right),
\end{align}
which satisfies $\mathcal C V \mathcal C^{-1}=-V^T$. The fields for heavy mesons are collected into
\begin{align}
P^{(c)}&=(D^0,D^+,D_s^+), \notag \\
P^{*(c)}&=(D^{*0},D^{*+},D^{*+}_{s}), \notag\\
P_1^{(c)}&=(D_1(2420)^0,D_1(2420)^+,D_{s1}(2536)^+), \notag\\
P_2^{(c)}&=(D_2(2460)^0,D_2(2460)^+,D_{s2}(2573)^+),\notag\\
P^{(b)}&=(B^+,B^0,B_s^0), \notag \\
P^{*(b)}&=(B^{*+},B^{*0},B^{*0}), \notag\\
P_1^{(b)}&=(B_1(5721)^+,B_1(5721)^0,B_{s1}(5830)^+), \notag\\
P_2^{(b)}&=(B_2(5747)^+,B_2(5747)^0,D_{s2}(5840)^+).
\end{align}
Note that $P_1$ and $P_2$ are $P$-wave mesons where the total angular momentum of light degrees of freedom is $s_l=3/2$. The $P$-wave mesons with $s_l=1/2$ will not be considered here because the charmed ones are too wide~\cite{ParticleDataGroup:2020ssz} to form bound states~\cite{Filin:2010se,Guo:2011dd} or considered as molecular states~\cite{Guo:2019dpg} and the bottomed ones have not been found experimentally.

\subsection{Coupling of light vector mesons and heavy baryons}

In the heavy quark limit, the ground states of heavy baryons $Qqq$ form an SU(3) antitriplet with $J^P = \frac12^+$ denoted by $B^{(Q)}_{\bar 3}$ and two degenerate sextets with $J^P = (\frac12,\frac32)^+$ denoted by $(B^{(Q)}_{6},B^{(Q)*}_6)$~\cite{Yan:1992gz},
\begin{align}
B^{(c)}_{\bar{3}}&=\left(\begin{array}{ccc}
0 & \Lambda_{c}^{+} & \Xi_{c}^{+} \\
-\Lambda_{c}^{+} & 0 & \Xi_{c}^{0} \\
-\Xi_{c}^{+} & -\Xi_{c}^{0} & 0
\end{array}\right),
\\ B^{(c)}_{6}&=\left(\begin{array}{ccc}
\Sigma_{c}^{++} & \frac{1}{\sqrt{2}} \Sigma_{c}^{+} & \frac{1}{\sqrt{2}} \Xi_{c}^{\prime+} \\
\frac{1}{\sqrt{2}} \Sigma_{c}^{+} & \Sigma_{c}^{0} & \frac{1}{\sqrt{2}} \Xi_{c}^{\prime 0} \\
\frac{1}{\sqrt{2}} \Xi_{c}^{\prime+} & \frac{1}{\sqrt{2}} \Xi_{c}^{\prime 0} & \Omega_{c}^{0}
\end{array}\right), \\
B_{6}^{(c)*}&=\left(\begin{array}{ccc}
\Sigma_{c}^{*++} & \frac{1}{\sqrt{2}} \Sigma_{c}^{*+} & \frac{1}{\sqrt{2}} \Xi_{c}^{*+} \\
\frac{1}{\sqrt{2}} \Sigma_{c}^{*+} & \Sigma_{c}^{* 0} & \frac{1}{\sqrt{2}} \Xi_{c}^{* 0} \\
\frac{1}{\sqrt{2}} \Xi_{c}^{*+} & \frac{1}{\sqrt{2}} \Xi_{c}^{* 0} & \Omega_{c}^{* 0}
\end{array}\right),\\
B^{(b)}_{\bar{3}}&=\left(\begin{array}{ccc}
0 & \Lambda_{b}^{0} & \Xi_{b}^{0} \\
-\Lambda_{b}^{0} & 0 & \Xi_{b}^{-} \\
-\Xi_{b}^{0} & -\Xi_{b}^{-} & 0
\end{array}\right),\\
 B^{(b)}_{6}&=\left(\begin{array}{ccc}
\Sigma_{b}^{+} & \frac{1}{\sqrt{2}} \Sigma_{b}^{0} & \frac{1}{\sqrt{2}} \Xi_{b}^{\prime0} \\
\frac{1}{\sqrt{2}} \Sigma_{b}^{0} & \Sigma_{b}^{-} & \frac{1}{\sqrt{2}} \Xi_{b}^{\prime -} \\
\frac{1}{\sqrt{2}} \Xi_{b}^{\prime0} & \frac{1}{\sqrt{2}} \Xi_{b}^{\prime -} & \Omega_{b}^{-}
\end{array}\right),\\
 B^{(b)*}_{6}&=\left(\begin{array}{ccc}
\Sigma_{b}^{*+} & \frac{1}{\sqrt{2}} \Sigma_{b}^{*0} & \frac{1}{\sqrt{2}} \Xi_{b}^{*0} \\
\frac{1}{\sqrt{2}} \Sigma_{b}^{*0} & \Sigma_{b}^{*-} & \frac{1}{\sqrt{2}} \Xi_{b}^{* -} \\
\frac{1}{\sqrt{2}} \Xi_{b}^{*0} & \frac{1}{\sqrt{2}} \Xi_{b}^{*-} & \Omega_{b}^{*-}
\end{array}\right).
\end{align}
Note that the heavy baryon multiplets have not been completely established experimentally and the experimental candidates in RPP~\cite{ParticleDataGroup:2020ssz} of the above baryons predicted by quark model are listed in Table~\ref{tab:baryonmass}. The isospin averaged masses will be used in this work and no isospin breaking will be considered. The $\Sigma_b^{(*)0}$, $\Xi_b^{\prime0}$ and $\Omega_b^{*0}$ have no experimental candidates, and thus we use $m_{\Sigma_b^{(*)}}=\lra{m_{\Sigma_b^{(*)+}}+m_{\Sigma_b^{(*)-}}}/2$ and $m_{\Xi_b'}=m_{\Xi_b^{\prime-}}$.

\begin{table}[t]
    \caption{The experimental candidates of heavy baryons predicted by quark model. The notations of experimental states are taken from RPP~\cite{ParticleDataGroup:2020ssz}. The $\Sigma_b^{(*)0}$, $\Xi_b^{\prime0}$ and $\Omega_b^{*0}$ do not have experimental candidates yet.} \label{tab:baryonmass}
    \centering
    \begin{ruledtabular}
    \begin{tabular}{cc|cc}
        model&  \ experimental\  & model& \ experimental\  \\
\hline
		$\Lambda_{c}^{+}$&$\Lambda_{c}^{+}$&$\Lambda_{b}^{0}$&$\Lambda_{b}^{0}$\\
\hline
		$\Xi_{c}^{+}$&$\Xi_{c}^{+}$&$\Xi_{b}^{0}$&$\Xi_{b}^{0}$ \\
\hline 
		$\Xi_{c}^{0}$&$\Xi_{c}^{0}$&$\Xi_{b}^{-}$&$\Xi_{b}^{-}$\\
\hline 
		$\Sigma_c$&$\Sigma_c(2455)$&$\Sigma_b$&$\Sigma_b$\\
\hline 
		$\Xi_{c}^{\prime+}$&$\Xi_{c}^{\prime+}$&$\Xi_{b}^{\prime0}$&$-$ \\
\hline 
		$\Xi_{c}^{\prime0}$&$\Xi_{c}^{\prime0}$&$\Xi_{b}^{\prime-}$&$\Xi_{b}^{\prime}(5935)^-$\\
\hline 	
    $\Omega_{c}^{0}$&$\Omega_{c}^{0}$& $\Omega_{b}^{-}$&$\Omega_{b}^{-}$\\
\hline 	
    $\Sigma_c^*$&$\Sigma_c(2520)$&$\Sigma_b^*$&$\Sigma_b^*$\\
\hline 
		$\Xi_{c}^{*+}$&$\Xi_{c}(2645)^+$&$\Xi_{b}^{*0}$&$\Xi_{b}(5945)^0$\\
\hline 
		$\Xi_{c}^{*0}$&$\Xi_{c}(2645)^0$&$\Xi_{b}^{*-}$&$\Xi_{b}(5955)^-$\\
\hline 	
		$\Omega_c^{*0}$&$\Omega_c(2770)^0$&$\Omega_b^{*0}$&$-$
    \end{tabular}
    \end{ruledtabular}
\end{table}

The Lagrangian for the coupling of heavy baryons and light mesons is constructed as~\cite{Liu:2011xc,Dong:2021juy} 
\begin{align}
\mathcal{L}_{BBV}&=\, i \beta_{B} \mathrm{tr}\left[\bar{B}^{(Q)}_{\bar{3}} v^{\mu}\left(\mathcal V_{\mu}-\rho_{\mu}\right) B^{(Q)}_{\bar{3}}\right]\notag\\
&-i \beta_{B}\mathrm{tr}\left[\bar{B}^{(\bar Q)}_{{3}} v^{\mu}\left(\mathcal V_{\mu}-\rho_{\mu}\right)^T B^{(\bar Q)}_{{3}}\right]\notag\\
&+ i \beta_{S} \operatorname{tr}\left[\bar{S}_{\nu}^{(Q)} v^{\mu}\left(\mathcal V_{\mu}-\rho_{\mu}\right) S^{(Q)\nu}\right]\notag\\
&-i \beta_{S}\operatorname{tr}\left[\bar{S}_{\nu}^{(\bar Q)} v^{\mu}\left(\mathcal V_{\mu}-\rho_{\mu}\right)^T S^{(\bar Q)\nu}\right],\label{eq:LSSV}
\end{align}
where the heavy baryons are expressed as bispinors,
\begin{align}
S^{(Q)}_{\mu}&=B_{6 \mu}^{(Q)*}- \frac{1}{\sqrt{3}}\left(\gamma_{\mu}+v_{\mu}\right) \gamma^{5} B^{(Q)}_{6},\\
\bar S^{(Q)}_{\mu}&=\bar B_{6 \mu}^{(Q)*}+ \frac{1}{\sqrt{3}}\bar B^{(Q)}_6\gamma^{5} \left(\gamma_{\mu}+v_{\mu}\right),
\end{align}
$v_\mu$ is the four-velocity of the heavy field, and tr denotes the traces over both the spinor and light flavor spaces.

\section{Molecular states from resonance-saturated constant interactions}\label{sec:3}

In the following we will solve the  Bethe-Salpeter equation $T=V+VGT$~\cite{Oller:1997ti} to search for poles of the scattering amplitude $T$. The interaction kernel (potential) $V$ is defined as $V=-\mathcal M$ with $\mathcal M$ the $2\to 2$ invariant scattering amplitude so that a negative $V$ means an attraction interaction. Such a potential is also the same as the nonrelativistic potential in the Schr\"odinger equation up to a mass factor. 

\begin{table}[tb]
    \caption{Values of the coupling parameters used in the calculations.} 
    \centering
    \begin{ruledtabular}
    \begin{tabular}{cccccc}
    $g_V$ &  $\beta$ & $\beta_2$ & $\zeta_1$ & $\beta_B$ & $\beta_S$ \\\hline
    5.8 & 0.9 & $-0.9$ & 0.16 & 0.87 & $-1.74$ \\
    \cite{Bando:1987br} & \cite{Isola:2003fh} & \cite{Dong:2019ofp} & \cite{Dong:2019ofp} & \cite{Liu:2011xc,Chen:2019asm} & \cite{Liu:2011xc,Chen:2019asm}
    \end{tabular}
    \end{ruledtabular}
    \label{tab:parameters}
\end{table}

In Table~\ref{tab:parameters}, we list the numerical values of the coupling constants used in this work with the corresponding references, which have been used in our previous work~\cite{Dong:2021juy}. Note that the signs of $\beta_B$ and $\beta_S$ adapted in this work are different from those in Ref.~\cite{Liu:2011xc}, the choice of which is in conflict with the molecular interpretation of the famous $P_c$ states as well as those obtained by flavor SU(4) relations~\cite{Wu:2010vk}.

\subsection{Potentials from light vector meson exchange}

The constant potentials of different systems assuming the saturation of the light vector meson exchange can be expressed uniformly as\footnote{ For the systems composed of two identical particles, such as $DD$, the symmetry factor $\frac12$ cancels the additional $u-$channel contribution and finally this equation holds valid.}
\begin{equation}
    V\approx -F \tilde \beta_1 \tilde \beta_2g_V^2\frac{2m_1m_2}{m_{\rm ex}^2},
    \label{eq:potential}
\end{equation}
with $m_1,m_2$ and $m_{\rm ex}$ the masses of the two heavy hadrons and the exchanged particle, respectively. $\tilde \beta_1$ and $\tilde \beta_2$ are the coupling constants for the two heavy hadrons with the vector mesons and are explicitly given as terms of the couplings in Eqs.~\eqref{eq:LPPV} and \eqref{eq:LSSV} as
\begin{itemize}
    \item $\tilde \beta_i=\beta$ for the $S$-wave charmed mesons,
    \item  $\tilde \beta_i=-\beta$ for the $P$-wave charmed mesons,
    \item  $\tilde \beta_i=\beta_B$ for the anti-triplet charmed baryons,
    \item  and $\tilde \beta_i=-\beta_S/2$ for the sextet charmed baryons.
\end{itemize}
$F$ is a group theory factor accounting for the light-flavor SU(3) information, and in our convention a positive $F$ means an attractive interaction. The values of $F$ for charmed-(anti)charmed and bottomed-(anti)bottomed systems are listed in Tables~\ref{tab:potentialsc}, \ref{tab:potentialsc1}, \ref{tab:potentialsb} and \ref{tab:potentialsb1} in Appendix~\ref{app:poten} for all combinations of heavy-(anti)heavy hadron pairs (the ones for the heavy-antiheavy systems have been given in our previous work~\cite{Dong:2021juy}). 
{Notice that we use the usual relativistic normalization for all the involved fields, i.e., a factor of $m_1 m_2$ has been multiplied to the amplitude derived from the Lagrangian in Eqs.~(\ref{eq:LPPV}, \ref{eq:LSSV}) so as to get $V$ in Eq.~\eqref{eq:potential}. }

It is important to note that the isoscalar meson exchange yields potentials with opposite signs for heavy-heavy and heavy-antiheavy systems while the isovector exchange leads to potentials with the same sign. Such observations are confirmed formally in Appendix~\ref{app:poten}.

\begin{table}[tb]
    \caption{Pole positions of double-charm-hadron systems with $I=0$ and $P=+$. $E_{\rm th}$ in the second column is the threshold in MeV. The results as given in the last columns corresponds to using the cutoff $\Lambda=0.5$ ($1.0$)~GeV for Eq.~(\ref{eq:GGF}) used to determine the subtraction constant $a(\mu)$ in Eq.~(\ref{eq:GDR}), respectively. In the last two columns, the first number in the parenthesis refers to the Riemann sheet (RS) where the pole is located while the second number means the distance between the pole position and the corresponding threshold, namely, $E_B \equiv E_{\rm th}-E_{\rm pole}$, in units of MeV.} \label{tab:pole0p}
    \centering
    \begin{ruledtabular}
    \begin{tabular}{l|cccc}

System & $E_{\rm th}$ [MeV] & $J^{P}$& \multicolumn{2}{c}{(RS, $E_B$ [MeV])} \\  & & & $0.5$ GeV & 1.0 GeV\\
\hline
$D D^{*}$&3876&$1^+$&(2, 3.58)&(1, 5.96)\\
$D^*D^{*}$&4017&$1^+$&(2, 2.68)&(1, 7.07)\\
$D_1D_1$&4844&$1^+$&(2, 0.321)&(1, 12.2)\\
$D_1D_2$&4885&$(1,2,3)^+$&(2, 0.277)&(1, 12.4)\\
$D_2D_2$&4926&$(1,3)^+$&(2, 0.237)&(1, 12.6)\\
\hline
$\Sigma_c{\Sigma}_c$&4907&$0^+$&(1, 2.72)&(1, 35.2)\\
$\Xi_c{\Xi}_c$&4939&$1^+$&(2, 43.4)&(2, 10.1)\\
$\Sigma_c{\Sigma}_c^*$&4972&$(1,2)^+$&(1, 2.79)&(1, 35.1)\\
$\Sigma_c^*{\Sigma}_c^*$&5036&$(0,2)^+$&(1, 2.86)&(1, 35.1)\\
$\Xi_c{\Xi}_c'$&5048&$(0,1)^+$&(2, 40.1)&(2, 8.55)\\
$\Xi_c{\Xi}_c^*$&5115&$(1,2)^+$&(2, 38.3)&(2, 7.73)\\
$\Xi_c'{\Xi}_c'$&5158&$1^+$&(2, 36.9)&(2, 7.14)\\
$\Xi_c^*{\Xi}_c'$&5225&$(1,2)^+$&(2, 35.2)&(2, 6.4)\\
$\Xi_c^*{\Xi}_c^*$&5292&$(1,3)^+$&(2, 33.4)&(2, 5.7)\\
\hline
${D}_1\Xi_c$&4891&$(\frac12,\frac32)^+$&(1, 2.78)&(1, 35.5)\\
${D}_2\Xi_c$&4932&$(\frac32,\frac52)^+$&(1, 2.83)&(1, 35.5)\\
${D}_1\Xi_c'$&5001&$(\frac12,\frac32)^+$&(1, 2.89)&(1, 35.4)\\
${D}_2\Xi_c'$&5042&$(\frac32,\frac52)^+$&(1, 2.94)&(1, 35.4)\\
${D}_1\Xi_c^{*}$&5068&$(\frac12,\frac32,\frac52)^+$&(1, 2.96)&(1, 35.3)\\
${D}_2\Xi_c^{*}$&5109&$(\frac12,\frac32,\frac52,\frac72)^+$&(1, 3.0)&(1, 35.3)\\
\hline
$D_{s1}{\Omega}_c$&5230&$(\frac12,\frac32)^+$&(1, 0.0298)&(1, 17.4)\\
$D_{s2}{\Omega}_c$&5264&$(\frac32,\frac52)^+$&(1, 0.039)&(1, 17.5)\\
$D_{s1}{\Omega}_c^*$&5301&$(\frac12,\frac32,\frac52)^+$&(1, 0.0474)&(1, 17.6)\\
$D_{s2}{\Omega}_c^*$&5335&$(\frac12,\frac32,\frac52,\frac72)^+$&(1, 0.0588)&(1, 17.7)\\
\end{tabular}
\end{ruledtabular}
\end{table}

\begin{table}[tb]
    \caption{Pole positions of double-charm-hadron systems with $I=0$ and $P=-$. See the caption for Table~\ref{tab:pole0p}.} \label{tab:pole0m}
    \centering
    \begin{ruledtabular}
    \begin{tabular}{l|cccc}
System & $E_{\rm th}$ [MeV] & $J^{P}$& \multicolumn{2}{c}{(RS, $E_B$ [MeV])} \\  & & & $0.5$ GeV & 1.0 GeV\\
\hline
$D\ {D}_1$&4289&$1^-$&(2, 2.48)&(1, 6.94)\\
$D\ {D}_2$&4330&$2^-$&(2, 1.65)&(1, 8.69)\\
$D^*{D}_1$&4431&$(0,1,2)^-$&(2, 1.35)&(1, 9.12)\\
$D^*{D}_2$&4472&$(1,2,3)^-$&(2, 1.0)&(1, 10.1)\\
\hline
${D}\ \Xi_c$&4337&$\frac12^-$&(1, 1.92)&(1, 35.3)\\
${D}\ \Xi_c'$&4446&$\frac12^-$&(1, 2.04)&(1, 35.4)\\
${D}^{*}\Xi_c$&4478&$(\frac12,\frac32)^-$&(1, 2.19)&(1, 35.5)\\
${D}\ \Xi_c^*$&4513&$\frac32^-$&(1, 2.11)&(1, 35.4)\\
${D}^{*}\Xi_c'$&4587&$(\frac12,\frac32)^-$&(1, 2.31)&(1, 35.5)\\
${D}^{*}\Xi_c^*$&4655&$(\frac12,\frac32,\frac52)^-$&(1, 2.38)&(1, 35.5)\\
\hline
$D_s{\Omega}_c$&4664&$\frac12^-$&(2, 0.168)&(1, 14.3)\\
$D_s{\Omega}_c^*$&4734&$\frac32^-$&(2, 0.129)&(1, 14.6)\\
$D_s^*{\Omega}_c'$&4807&$(\frac12,\frac32)^-$&(2, 0.0507)&(1, 15.3)\\
$D_s^*{\Omega}_c^*$&4878&$(\frac12,\frac32,\frac52)^-$&(2, 0.0308)&(1, 15.6)\\
\end{tabular}
\end{ruledtabular}
\end{table}

\begin{table}[tb]
    \caption{Pole positions of double-charm-hadron systems with $I=1/2$ and $P=-$. See the caption for Table~\ref{tab:pole0p}.} \label{tab:pole05m}
    \centering
    \begin{ruledtabular}
    \begin{tabular}{l|cccc}
System & $E_{\rm th}$ [MeV] & $J^{P}$& \multicolumn{2}{c}{(RS, $E_B$ [MeV])} \\  & & & $0.5$ GeV & 1.0 GeV\\
\hline
${D}\ \Lambda_c$&4154&$\frac12^-$&(2, 3.44)&(1, 5.62)\\
${D}^{*}\Lambda_c$&4295&$(\frac12,\frac32)^-$&(2, 2.53)&(1, 6.73)\\
${D}\ \Sigma_c$&4321&$\frac12^-$&(1, 5.81)&(1, 50.5)\\
${D}\ \Sigma_c^{*}$&4385&$\frac32^-$&(1, 5.85)&(1, 50.2)\\
${D}^{*}\Sigma_c$&4462&$(\frac12,\frac32)^-$&(1, 5.97)&(1, 49.7)\\
${D}^{*}\Sigma_c^{*}$&4527&$(\frac12,\frac32,\frac52)^-$&(1, 6.01)&(1, 49.5)\\
\hline
$D_s{\Xi}_c$&4438&$\frac12^-$&(2, 25.7)&(2, 1.76)\\
$D_s{\Xi}_c'$&4547&$\frac12^-$&(2, 23.7)&(2, 1.29)\\
$D_s^*{\Xi}_c$&4582&$(\frac12,\frac32)^-$&(2, 21.8)&(2, 0.882)\\
$D_s{\Xi}_c^*$&4614&$\frac32^-$&(2, 22.6)&(2, 1.05)\\
$D_s^*{\Xi}_c'$&4691&$(\frac12,\frac32)^-$&(2, 20.0)&(2, 0.564)\\
$D_s^*{\Xi}_c^*$&4758&$(\frac12,\frac32,\frac52)^-$&(2, 19.0)&(2, 0.416)\\
\end{tabular}
\end{ruledtabular}
\end{table}

\begin{table}[tb]
    \caption{Pole positions of double-charm-hadron systems with $I=1/2$ and $P=+$. See the caption for Table~\ref{tab:pole0p}.} \label{tab:pole05p}
    \centering
    \begin{ruledtabular}
    \begin{tabular}{l|cccc}
System & $E_{\rm th}$ [MeV] & $J^{P}$& \multicolumn{2}{c}{(RS, $E_B$ [MeV])} \\  & & & $0.5$ GeV & 1.0 GeV\\
\hline
${D}_1\Lambda_c$&4708&$(\frac12,\frac32)^+$&(2, 1.04)&(1, 9.31)\\
${D}_2\Lambda_c$&4750&$(\frac32,\frac52)^+$&(2, 0.95)&(1, 9.51)\\
${D}_1\Sigma_c$&4876&$(\frac12,\frac32)^+$&(1, 6.25)&(1, 47.5)\\
${D}_2\Sigma_c$&4917&$(\frac32,\frac52)^+$&(1, 6.27)&(1, 47.3)\\
${D}_1\Sigma_c^{*}$&4940&$(\frac12,\frac32,\frac52)^+$&(1, 6.28)&(1, 47.2)\\
${D}_2\Sigma_c^{*}$&4981&$(\frac12,\frac32,\frac52,\frac72)^+$&(1, 6.29)&(1, 47.0)\\
\hline
$D_{s1}{\Xi}_c$&5005&$(\frac12,\frac32)^+$&(2, 14.2)&(2, 0.00911)\\
$D_{s2}{\Xi}_c$&5039&$(\frac32,\frac52)^+$&(2, 13.8)&(2, 0.00176)\\
$D_{s1}{\Xi}_c'$&5114&$(\frac12,\frac32)^+$&(2, 12.8)&(1, 0.00636)\\
$D_{s2}{\Xi}_c'$&5148&$(\frac32,\frac52)^+$&(2, 12.4)&(1, 0.0175)\\
$D_{s1}{\Xi}_c^*$&5181&$(\frac12,\frac32,\frac52)^+$&(2, 12.0)&(1, 0.0319)\\
$D_{s2}{\Xi}_c^*$&5215&$(\frac12,\frac32,\frac52,\frac72)^+$&(2, 11.6)&(1, 0.0532)\\
\end{tabular}
\end{ruledtabular}
\end{table}

\subsection{Poles of molecular states}

Given the constant interactions between a pair of heavy-heavy hadrons, we solve the single channel BS equation that is factorized into an algebraic equation,
\begin{align}
    T=\frac{V}{1-VG},
\end{align}
to search for poles of the scattering amplitude and in turn to give a rough but overall picture of the spectrum of possible molecular states. Here $G$ is the one loop two-body propagator that after the dimensional regularization (DR)~\cite{Veltman:1994wz} reads\footnote{There are typos in the expression of $G(E)$ in the published version of the previous work~\cite{Dong:2021juy}.}
\begin{align}
    G(E)=&\frac1{16\pi^2}\bigg\{a(\mu)+\log\frac{m_{1}^2}{\mu^2}+\frac{m_{2}^2-m_{1}^2+s}{2s} \log\frac{m_{2}^2}{m_{1}^2} \nonumber\\
&+\frac{k}{E} \Big[ 
\log\left(2k E+s+\Delta\right) + 
\log\left(2k E+s-\Delta\right) \nonumber\\ 
& -  
\log\left(2k E-s+\Delta\right) - 
\log\left(2k E-s-\Delta\right)
\Big]\bigg\},\label{eq:GDR}
\end{align}
where $s=E^2$, $m_{1}$ and $m_{2}$ are the particle masses, $\Delta= m_{1}^2-m_{2}^2$, and
\begin{align}
k=\frac1{2E}\lambda^{1/2}(E^2,m_{1}^2,m_{2}^2)\label{eq:3-momentum}
\end{align}
is the corresponding three-momentum magnitude with $\lambda(x,y,z)=x^2+y^2+z^2 - 2xy - 2yz - 2xz$ for the K\"all\'en triangle function. The DR scale $\mu$ is chosen to be 1~GeV and its variation can be absorbed into the subtraction constant $a(\mu)$. For the single channel case considered in our paper, there are two Riemann sheets (RSs) that are defined as Im$(k)>0$ on the first RS while Im$(k)<0$ on the second RS.

Another way to regularize the loop integral is to introduce a Gaussian form factor, namely\footnote{ Note that unlike the dimensional regularization, this Gaussian form factor will violate unitarity, i.e. ${\rm Im}\left[T^{-1}(E)\right]=-i\rho(E)$ with $\rho(E)$ the two body phase space factor, if the same constant contact term is used. However, if we focus on the near threshold bound or virtual states, such a difference between Eq.~(\ref{eq:GDR}) and Eq.~(\ref{eq:GGF}) is negligible.},  
\begin{align}
G(E) =&\, \int \frac{l^2 dl}{4\pi^2} \frac{\omega_1+\omega_2}{\omega_1\omega_2} 
\frac{e^{-2l^2/\Lambda^2} }{E^2-(\omega_1+\omega_2)^2+i\epsilon} ,\label{eq:GGF}
\end{align}
with $\omega_i=\sqrt{m_i^2+l^2}$. The cutoff $\Lambda$ is chosen in the range of $0.5\sim 1.0$~GeV, which is believed phenomenologically adequate~\cite{Epelbaum:2008ga,Nieves:2012tt,Guo:2013sya}, and then the subtraction constant $a(\mu)$ in DR is determined by matching the values of $G$ from these two methods at threshold.  In the following we will use the DR loop with the so-determined subtraction constant for numerical calculations.

\begin{figure*}
    \centering
    \includegraphics[width=0.46\linewidth]{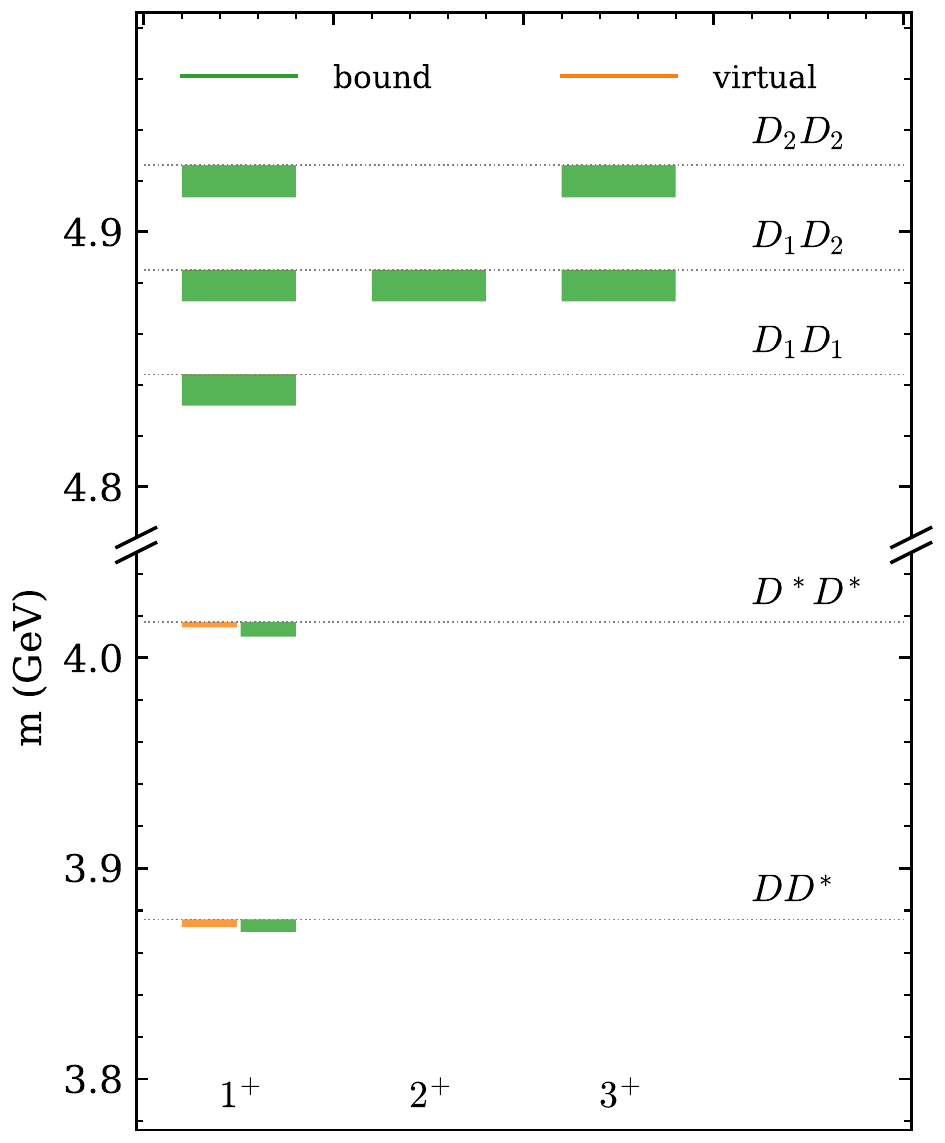}
    \includegraphics[width=0.53\linewidth]{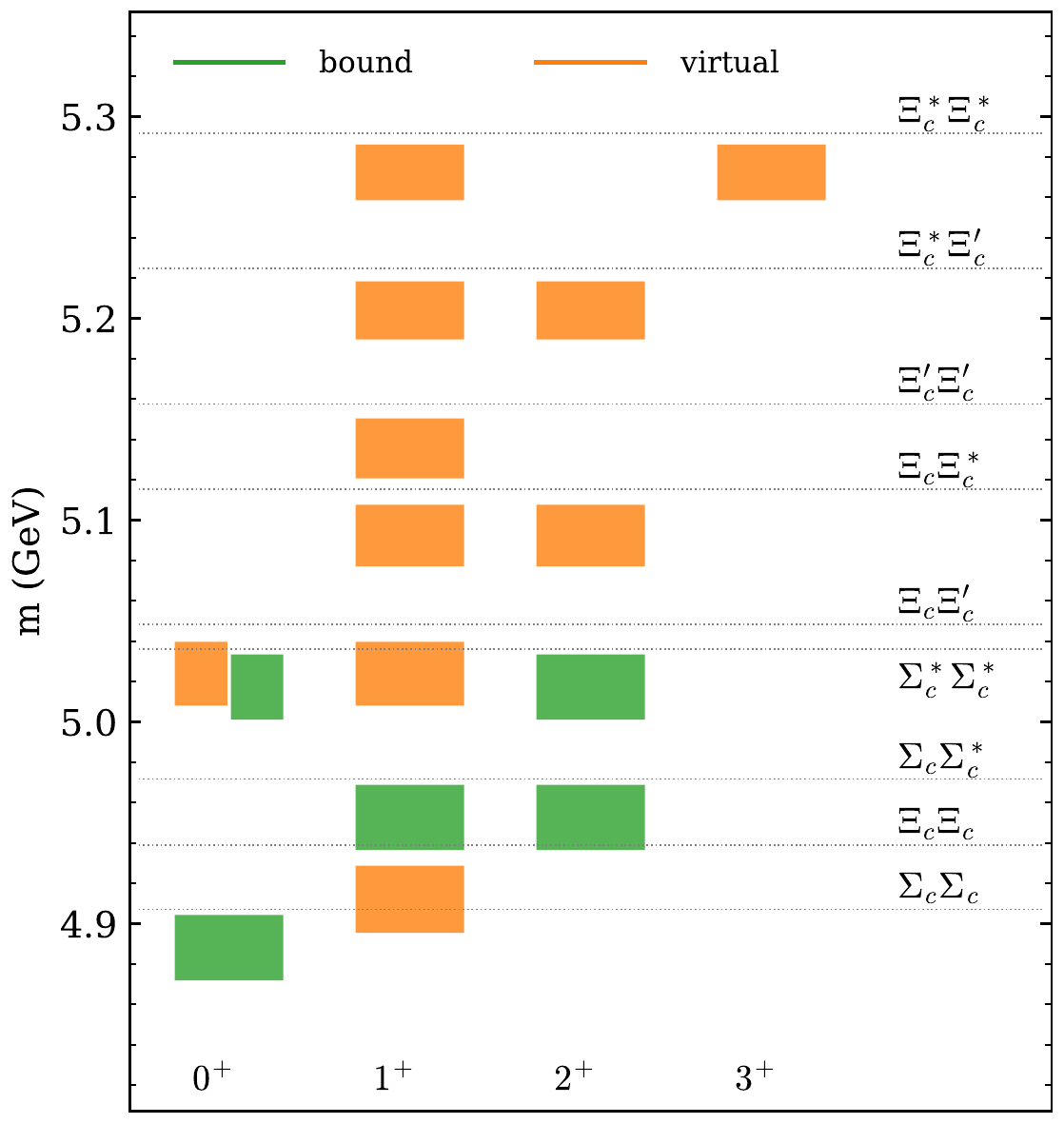}
    \caption{The spectrum of hadronic molecules consisting of a pair of charmed mesons or baryons with $I=0$ and $P=+$. The colored rectangle, green for a bound state and orange for a virtual state, covers the range of the pole position for a given system with the cutoff $\Lambda$ varying in the range of $[0.5, 1.0]$~GeV.Thresholds are marked by dotted horizontal lines. The rectangle closest to, but below, the threshold corresponds to the hadronic molecule in that system.  { In some cases, e.g., $DD^*$,  there are two rectangles for one system, with the upper edges exactly at the threshold. This corresponds to the situation that the pole moves from the second RS (left orange) to the first RS (right green) when $\Lambda$ increases in the considered range.} In some {other} cases where the pole positions of two systems overlap, small rectangles are used with the left (right) one for the system with the higher (lower) threshold. }\label{fig:specBB0}
\end{figure*}

\begin{figure}
    \centering
    \includegraphics[width=\linewidth]{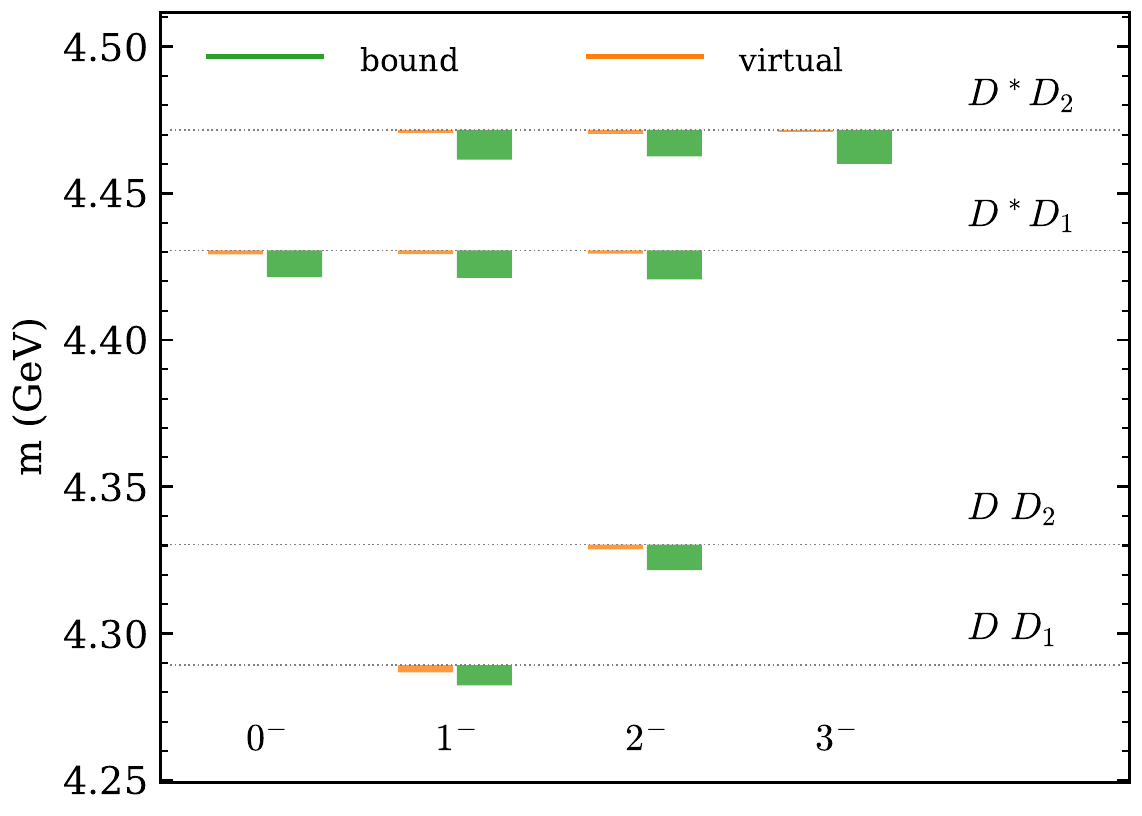}
    \caption{The spectrum of hadronic molecules consisting of a pair of charmed mesons with $I=0$ and $P=-$. See the caption for Fig.~\ref{fig:specBB0}. }\label{fig:specDD0}
\end{figure}

\begin{figure*}
    \centering
    \includegraphics[width=0.48\linewidth]{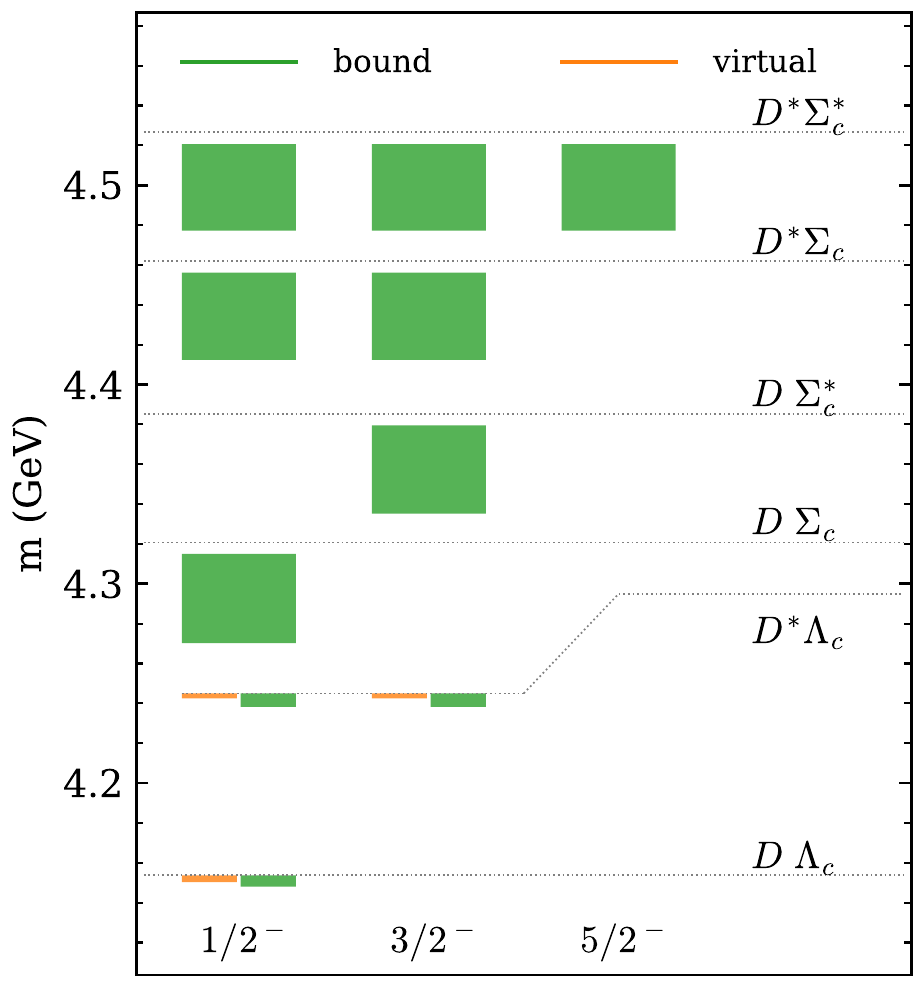}
    \includegraphics[width=0.48\linewidth]{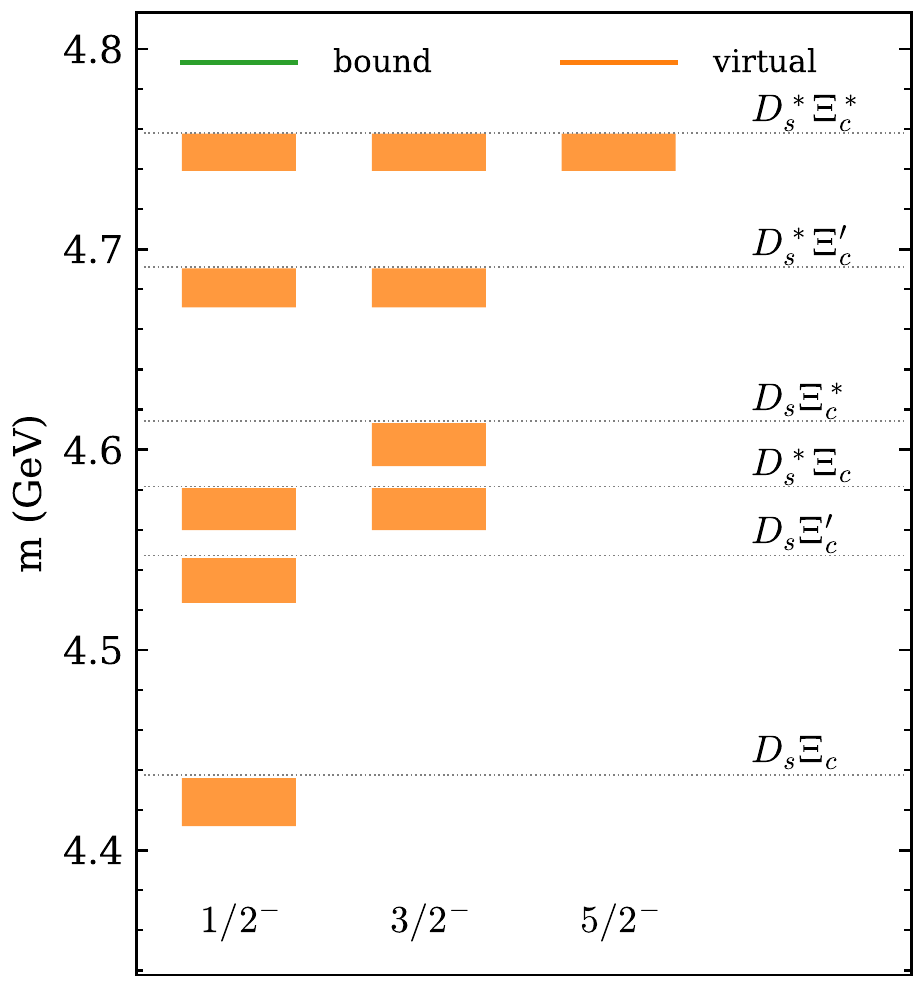}    \caption{The spectrum of hadronic molecules consisting of a pair of charmed meson and charmed baryon with $I=1/2$ and $P=-$. See the caption for Fig.~\ref{fig:specBB0}. The right part of the dashed line for $D^*\Lambda_c$ marks the real threshold while the left part is deformed to avoid being covered by the rectangle of $D\Sigma_c$ system.}\label{fig:specDBm5}
\end{figure*}

\begin{figure*}
    \centering
    \includegraphics[width=0.48\linewidth]{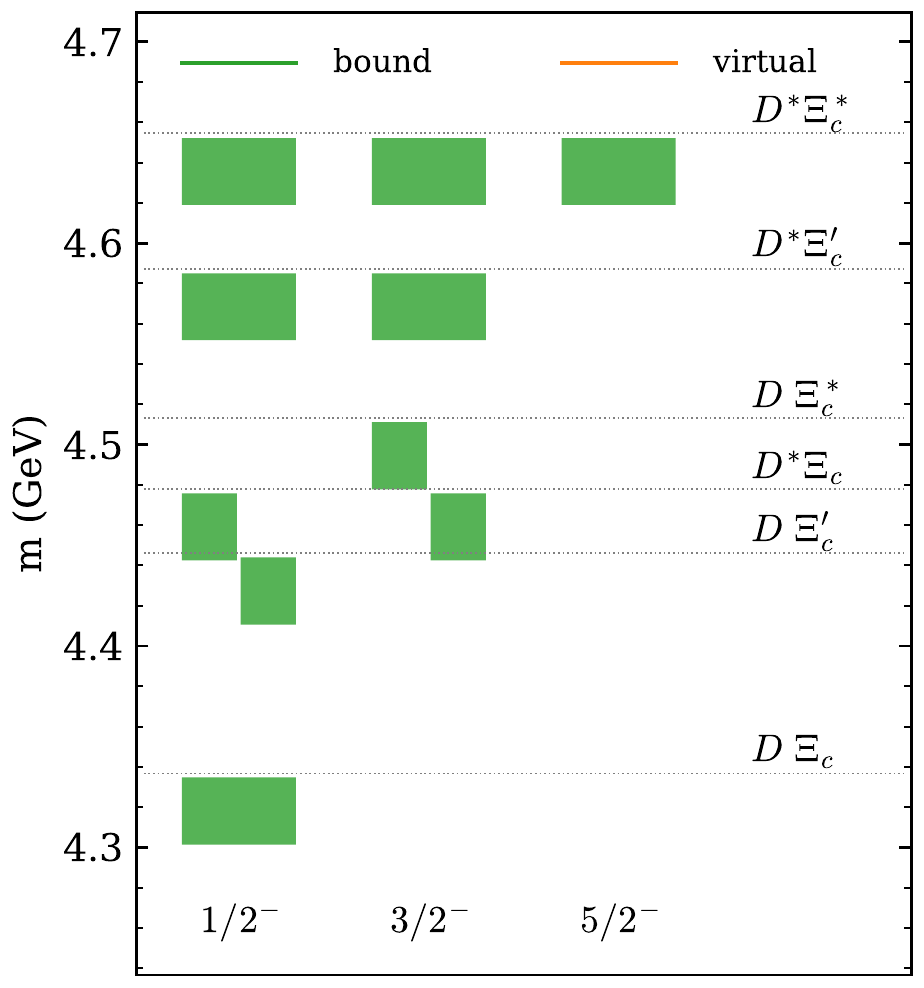}
    \includegraphics[width=0.48\linewidth]{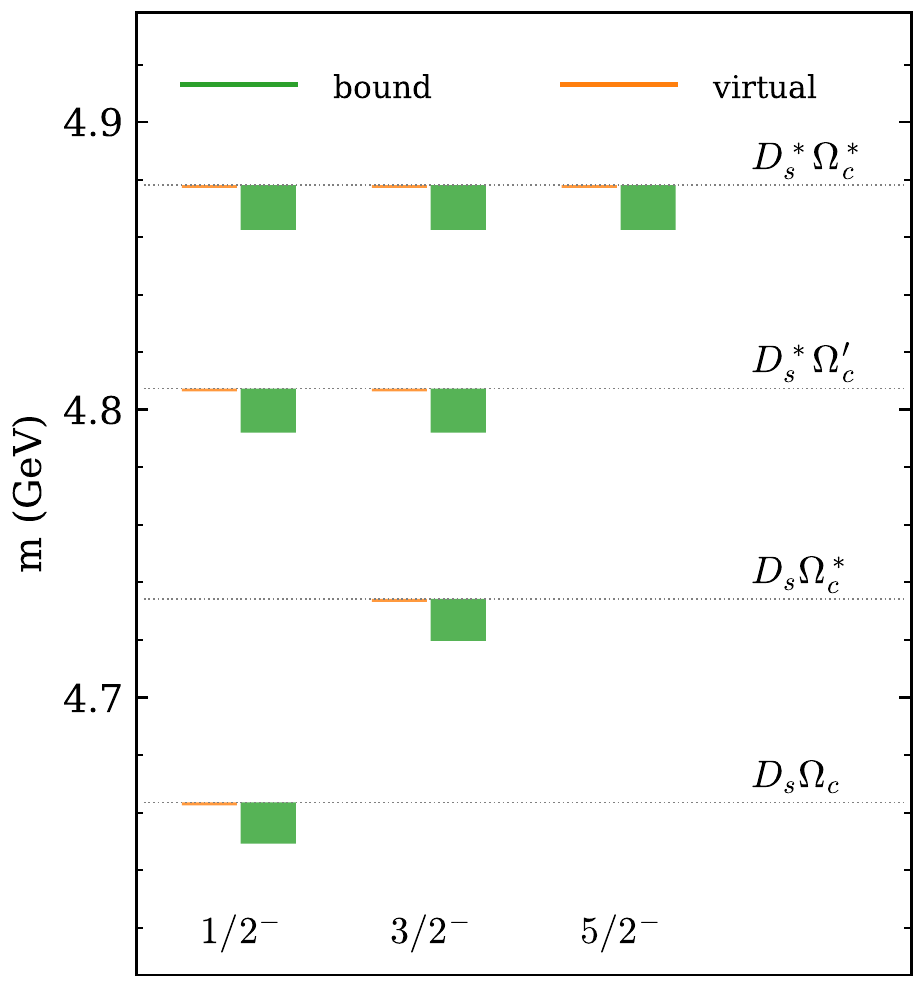}
    \caption{The spectrum of hadronic molecules consisting of a pair of charmed meson and charmed baryon with $I=0$ and $P=-$. See the caption for Fig.~\ref{fig:specBB0}. }\label{fig:specDBm0}
\end{figure*}

\begin{figure*}
    \centering
    \includegraphics[width=0.48\linewidth]{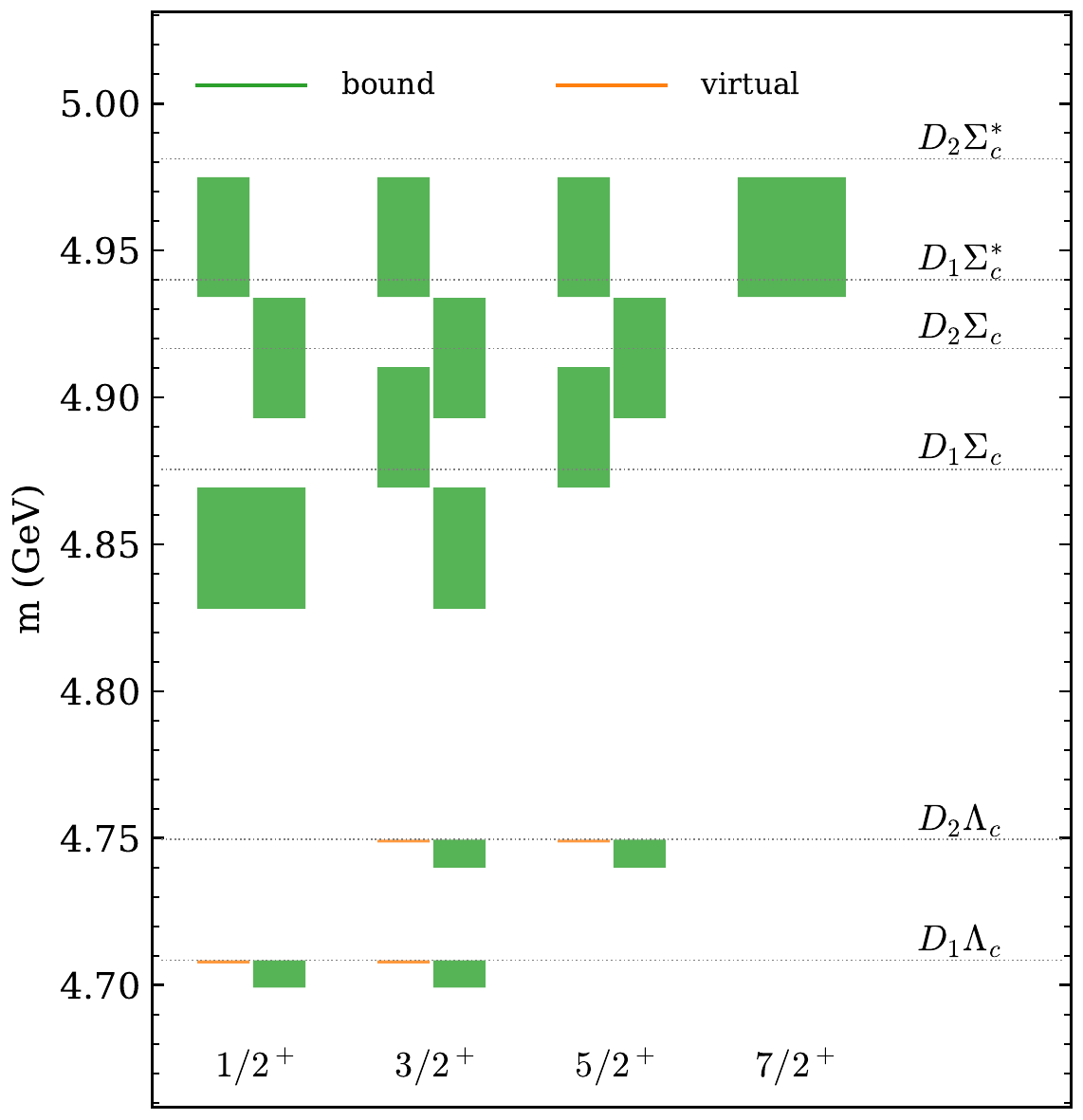}
    \includegraphics[width=0.48\linewidth]{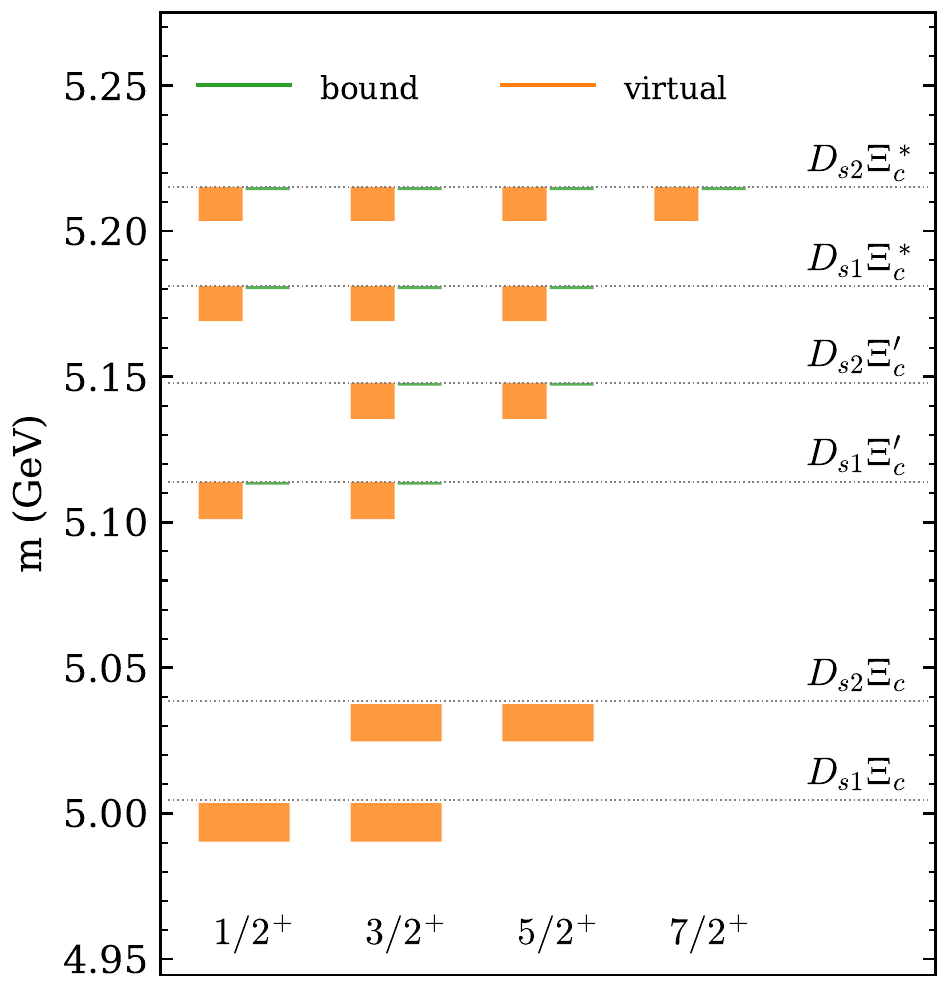}    \caption{The spectrum of hadronic molecules consisting of a pair of charmed meson and charmed baryon with $I=1/2$ and $P=+$. See the caption for Fig.~\ref{fig:specBB0}. }\label{fig:specDBp5}
\end{figure*}

\begin{figure*}
    \centering
    \includegraphics[width=0.48\linewidth]{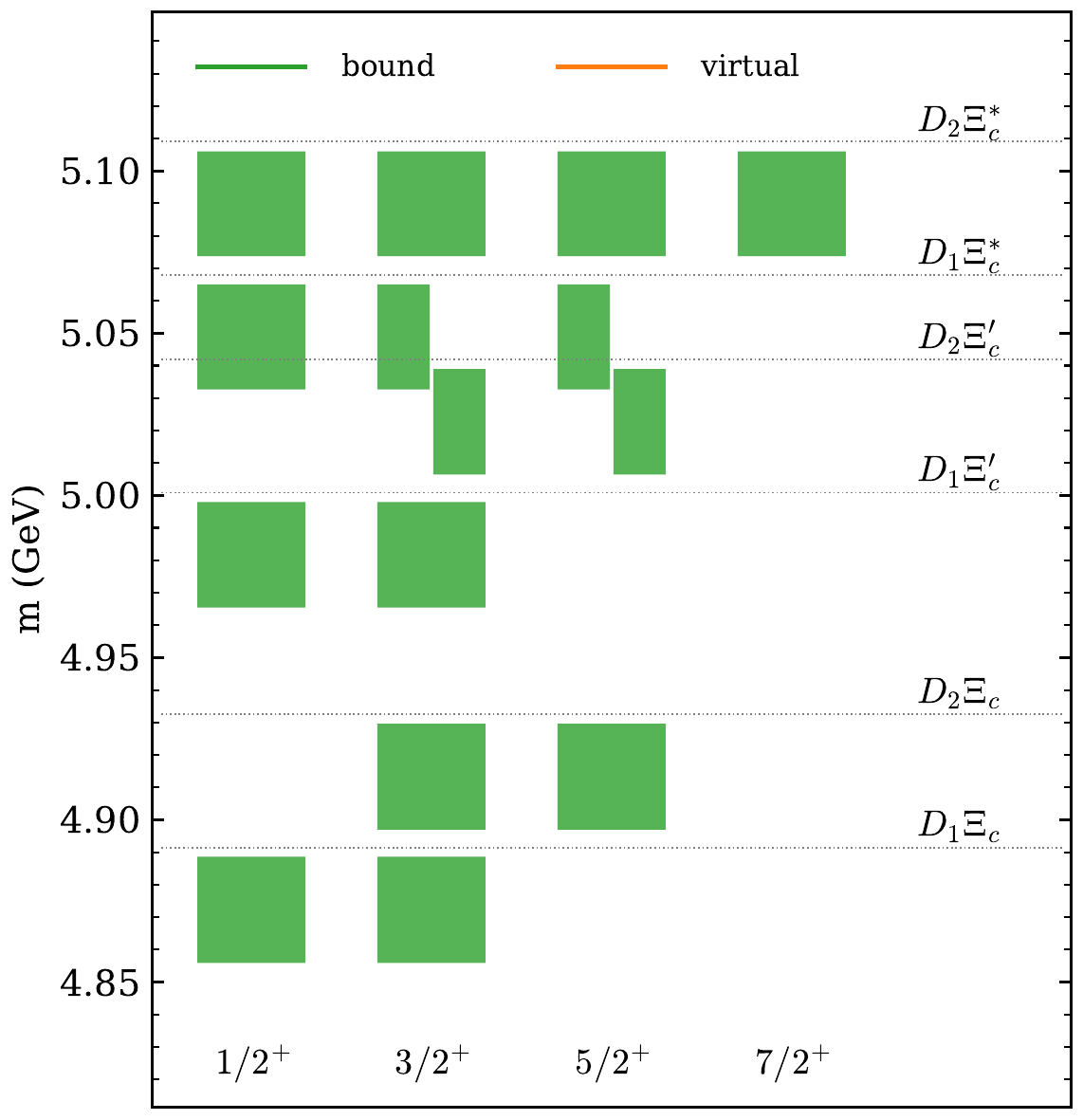}
    \includegraphics[width=0.48\linewidth]{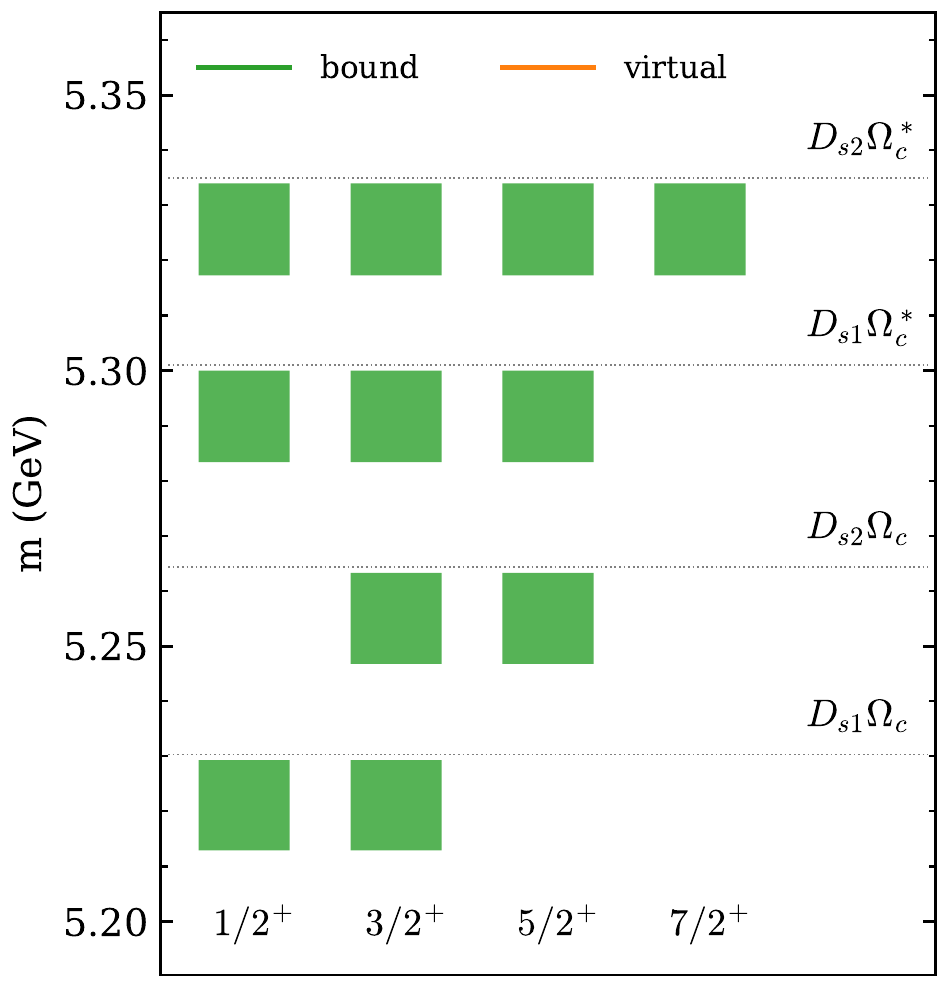}
    \caption{The spectrum of hadronic molecules consisting of a pair of charmed meson and charmed baryon with $I=0$ and $P=+$. See the caption for Fig.~\ref{fig:specBB0}. }\label{fig:specDBp0}
\end{figure*}

In Tables~\ref{tab:pole0p}, \ref{tab:pole0m}, \ref{tab:pole05m} and \ref{tab:pole05p}, we list all the pole positions of the double-charm-hadron systems which have attractive interactions, corresponding to the masses of hadronic molecules (bound states on 1st RS or virtual states on 2nd RS). For better illustration, these states are also shown in Figs.~\ref{fig:specBB0}, \ref{fig:specDD0}, \ref{fig:specDBm5}, \ref{fig:specDBm0}, \ref{fig:specDBp5} and \ref{fig:specDBp0} together with the corresponding thresholds. Considering the constant contact interactions saturated by the light vector meson exchange with the coupled-channel effects neglected, we obtain a spectrum of 124 hadronic molecules in total. At least the same number of molecules are expected to exist for each of the charm-bottom and bottom-bottom systems since it is easier to form a bound state with the same attraction strength due to the heavier reduced masses; there could be even more as if the ground state is deeply bound excited states might exist as well as illustrated in the J\"ulich meson-exchange model for hidden-bottom pentaquark-like hadronic molecules in Ref.~\cite{Shen:2017ayv}.

{ Besides the predicted molecular states mentioned above, where only $\omega, \rho$ and $\phi$ exchanges are considered, we need pay some extra attentions to the $K^*$ exchange, which is absent in the heavy-antiheavy systems due to the sizeable symmetry breaking between $s$ quark and $u/d$ quark. In most cases, $K^*$ exchange, if allowed, will contribute a repulsive potential, see Table~\ref{tab:potentialsc}, unless two channels are close enough and can be linearly combined to an isospin-like (called $U/V-$spin in Ref.~\cite{Meng:2020ihj}) eigenstates. Explicitly, the thresholds of $DD^*_s$ and $D^*D_s$ are close enough and they may form two $U/V-$spin eigenstates,
\begin{align}
    U/V=\pm: DD_s^*\pm D^*D_s
\end{align}
The potentials of $K^*$ exchange are repulsive and attractive, respectively, the strength of which can be also described by Eq.~(\ref{eq:potential}) with $|F|=1$ and $m_{\rm ex}=m_{K^*}$. Therefore, we expect a near threshold virtual or bound states in the $U/V=-1$ systems of $DD_s^*$. Some other similar systems, such as $\Sigma_c\Xi^*_c$ and $\Sigma_c^*\Xi_c$, the thresholds are not that close and more efforts should be paid to the coupled channel efforts, which is beyond the scope of this work.
}

\section{Discussions of selected systems}\label{sec:4}

\subsection{Heavy meson-meson molecules v.s. doubly heavy tetraquarks}\label{subsec:4A}

From Table~\ref{tab:potentialsc}, one sees that the attractions strength for the isoscalar $D^{(*)}D^*$ is half of that for the isoscalar $C=+$ $D^{(*)}\bar D^*$ pairs.\footnote{For the isoscalar systems, $V_{D^{(*)}D^*}=\frac32V_\rho-\frac12V_\omega$, which is about half of $V_{D^{(*)}\bar D^*}=\frac32V_\rho+\frac12V_\omega$ considering $V_\rho\approx V_\omega$, where $V_\rho$ and $V_\omega$ are the potentials due to the exchange of $\rho$ and $\omega$, respectively.} 
Thus, for the $X(3872)$ as a $D\bar D^*$ bound state with $J^{PC}=1^{++}$,\footnote{Notice that the narrow structure $X(4014)$ reported recently by Belle~\cite{Belle:2021nuv} is an excellent candidate for the $2^{++}$ $D^*\bar D^*$ molecule~\cite{Tornqvist:1993ng,Molina:2009ct,Nieves:2012tt,Guo:2013sya,Albaladejo:2015dsa,Baru:2016iwj,Dong:2021juy}, and its mass $(4014.4\pm4.1\pm0.5)$~MeV and width $(6\pm16\pm6)$~MeV nicely agree with the predictions in Ref.~\cite{Albaladejo:2015dsa}.
The $D\bar D$ bound state obtained in the same way~\cite{Dong:2021juy} (for early predictions, see Refs.~\cite{Zhang:2006ix,Gamermann:2006nm,Liu:2008tn,Wong:2003xk,Nieves:2012tt,HidalgoDuque:2012pq}) also receives support from a recent lattice QCD calculation~\cite{Prelovsek:2020eiw}.
} the $D^{(*)}D^*$ system would be less.
From Fig.~\ref{fig:specDD0} we see indeed that the isoscalar $D^{(*)}D^{*}$ systems are at the edge of forming bound states. While for the isovector ones, the potential from $\rho$ and $\omega$ exchange is repulsive and thus no molecules are expected. The situations for $D^{(*)}D_{(1,2)}$ and $D_{(1,2)}D_{(1,2)}$ systems are all similar. 

The $T_{cc}^+$ state was recently observed in the invariant mass distribution of $D^0D^0\pi^+$ by LHCb~\cite{LHCb:2021vvq,LHCb:2021auc}. { The pole from an analysis using a unitarized Breit-Wigner parameterization, considering the momentum-dependent width of the $D^*$ from its decays, is located in the complex energy plane at 
\begin{align}
    -360 \pm40 ^{+4}_{-0} - i\,(24\pm2 ^{+0}_{-7})~{\rm keV},
\end{align}
with the real part defined as the energy relative to the $D^0D^{*+}$ threshold~\cite{LHCb:2021auc}.}
\footnote{{A simple Breit-Wigner parametrization of the peak convolved with energy resolution leads to a width of $(410\pm165)$~keV~\cite{LHCb:2021vvq}.} Note that the $T_{cc}^+$ is very close to the $D^0D^{*+}$ threshold, and thus the Flatt\'e parameterization is more proper than the Breit-Wigner one; the former leads to a asymmetric line shape automatically while the latter does not. It is also worthwhile to notice that the width of a Flatt\'e line shape can be much smaller than that of the Breit-Wigner one when convolved with the energy resolution, as nicely illustrated in the careful analysis of the $X(3872)$ line shape by the LHCb Collaboration~\cite{LHCb:2020xds}. } This new state is a good candidate of the isoscalar $DD^*$ molecule, in agreement with our result. Before this experiment signal, lots of predictions of such doubly charmed tetraquark states, being either hadronic molecules~\cite{Janc:2004qn,Ohkoda:2012hv,Li:2012ss,Liu:2019stu,Liu:2020nil,Yang:2009zzp} or compact tetraquark states~\cite{Carlson:1987hh,Silvestre-Brac:1993zem,Silvestre-Brac:1993wyf,Semay:1994ht,Gelman:2002wf,Vijande:2003ki,Navarra:2007yw,Ebert:2007rn,Vijande:2007rf,Lee:2009rt,Abud:2009rk,Yang:2009zzp,Karliner:2013dqa,Feng:2013kea,Luo:2017eub,Karliner:2017qjm,Eichten:2017ffp,Wang:2017uld,Hyodo:2017hue,Cheung:2017tnt,Park:2018wjk,Junnarkar:2018twb,Deng:2018kly,Yang:2019itm,Tan:2020ldi,Lu:2020rog,Braaten:2020nwp,Gao:2020ogo,Cheng:2020wxa,Noh:2021lqs,Faustov:2021hjs}, have been made with various methods.

\subsubsection{Heavy meson-meson molecules}

The doubly heavy tetraquark states in molecular configurations were widely investigated in the literature~\cite{Carlson:1987hh,Richards:1990xf,Manohar:1992nd,Tornqvist:1993ng,Mihaly:1996ue,Stewart:1998hk,Barnes:1999hs,Janc:2004qn,Yang:2009zzp,Molina:2010tx,Carames:2011zz,Ohkoda:2012hv,Li:2012ss,Sun:2012sy,Sun:2012sy,Xu:2017tsr,Carames:2018tpe,Michael:1999nq,Pennanen:1999xi,Wagner:2011ev,Guerrieri:2014nxa,Ikeda:2013vwa,Detmold:2007wk,Bali:2011gq,Brown:2012tm,Bicudo:2012qt,Bicudo:2015vta,Bicudo:2015kna,Bicudo:2012qt,Brown:2012tm,Bicudo:2015vta,Bicudo:2015kna,Bicudo:2016ooe,Karliner:2015ina,SanchezSanchez:2017xtl,Sakai:2017avl,Wang:2018atz,Yu:2019sxx,Liu:2019stu,Ding:2020dio,Liu:2020nil,Meng:2020knc,Ding:2021igr} where predictions of many doubly heavy tetraquark states including the $T_{cc}$ mentioned above as well as other systems were made. 
 
The interaction between a pair of heavy mesons was estimated using the Born-Oppenheimer approximation in the MIT bag model in Ref.~\cite{Carlson:1987hh}. The $I(J^P)=0(1^+)$ di-meson $T(bb\bar q\bar q)$ was found to be a bound state about 70~MeV below the $BB^*$ threshold while the situations of $T(bc\bar q\bar q)$ and $T(cc\bar q\bar q)$ were uncertain.
 
 By solving the double-charm tetraquark system with two realistic potential models, it was found in Ref.~\cite{Janc:2004qn} that the ground state tetraquark state has a configuration of $DD^*$ molecule. In the constituent quark model, it was also found that the di-meson configurations of $QQ\bar q\bar q$ can be bound~\cite{Yang:2009zzp}.

In Ref.~\cite{Yu:2019sxx}, the $B^{(*)}B^{(*)}$ scattering amplitude was explored in the constituent interchange model and the $I(J^P)=0(1^+)$ $BB^*$ and $B^*B^*$ bound states together with virtual states with some other quantum numbers were found.

 In Ref.~\cite{Manohar:1992nd}, it was found that the long-range potential from one-pion exchange may be attractive enough to bind the $BB^*$ system but insufficient for the $DD^*$ system to bind, which has a smaller reduced mass. The existence of $DD^*$ was also disfavored by the one pion exchange in Ref.~\cite{Tornqvist:1993ng}. While the one-boson exchange model predicts that the isoscalar $DD^*$ system can form an $S$-wave bound state with a binding energy of $62.3$~MeV (depending on the cutoff) with the $\pi,\rho$ and $\omega$ exchanges~\cite{Ohkoda:2012hv}, $3\sim40$ MeV with the $\pi,\rho$ and $\omega$ exchanges~\cite{Liu:2020nil}, $0.47\sim 43$~MeV (depending on the cutoff) after including the $\pi,\sigma,\rho$ and $\omega$ exchanges and coupled channel effects~\cite{Li:2012ss} or $3^{+15}_{-~4}$~MeV without coupled channels but with the cutoff fixed by producing the correct binding energy of $X(3872)$~\cite{Liu:2019stu}. Besides, some other doubly heavy molecules including $D_{(s)}^{(*)} D_{(s)}^{(*)}$, $\bar{B}_{(s)}^{*} \bar{B}_{(s)}^{(*)}$ and $D^{(*)} \bar{B}^{(*)}$ with different quantum numbers are predicted in these works as well as in Ref.~\cite{Ding:2021igr}.  In Ref.~\cite{Xu:2017tsr}, the potential between $DD^*$ from the one-pion exchange  supplemented by the contact term and the exchange of two pions was investigated in a chiral effective field theory, and an isovector state was found to be bound while the scalar one was not.
 
 With the potential from a chiral constituent quark model, the Lippmann–Schwinger equation was solved for the $DD$-$DD^*$-$D^*D^*$ coupled channels in Ref.~\cite{Carames:2011zz} and a stable doubly charmed meson with $I(J^P)=0(1^+)$ was predicted. In Ref.~\cite{Carames:2018tpe}, the similar strategy yielded isoscalar $0^+$ and $1^+$ $bc\bar q\bar q$ bound states, which are stable against strong interaction, but the isovector systems were found unbound. 

 We can notice that the results from these works, as well as those obtained here, do not all agree with each other. There are at least two possible reasons: the form of the potential, in particular the treatment of the short-distance part, is different; the parameter values are different. In the spirit of effective field theory, different treatments of the short-distance potential correspond to taking different values for the contact terms. 
 In any case, most of the literature tends to agree that it is easier for the isoscalar $DD^*/BB^*$ to bind than the isovector combinations, as is the conclusion in our paper as well. 
 Our model assumes the short-distance contact terms are saturated by the light-vector-meson exchange, and the long- or mid-range attractive potential from the $\pi$ or $\sigma$ exchanges and the coupled channel effects may change the hadronic molecule spectrum quantitatively to some extent.

There are also lattice calculations of the potential between a pair of heavy mesons~\cite{Richards:1990xf,Mihaly:1996ue,Stewart:1998hk,Wagner:2011ev,Guerrieri:2014nxa}. The $DD^{(*)}$ interactions were calculated on lattice~\cite{Ikeda:2013vwa} and it was found that the potentials of the isovector systems are repulsive while those of the isoscalar systems are attractive, qualitatively in line with the contact interactions from the vector meson exchange reported here (see Table~\ref{tab:potentialsc}). 
The isoscalar $B^{(*)}B^{(*)}$ interaction from  lattice calculations~\cite{Pennanen:1999xi,Detmold:2007wk,Bali:2011gq,Brown:2012tm,Bicudo:2012qt,Bicudo:2015vta,Bicudo:2015kna} in the static $b$ quark limit is attractive in the short range while for the isovector one, the attraction is weaker. It was found plausible for $\bar b\bar b qq$ di-mesons to be stable under strong interactions~\cite{Michael:1999nq}. In Refs.~\cite{Bicudo:2012qt,Brown:2012tm,Bicudo:2015vta,Bicudo:2015kna,Bicudo:2016ooe}, the Schr\"odinger equation with the obtained potential yields results that the $I(J^P)=0(1^+)$ $\bar b\bar b qq$ system has an attractive potential between two $B^{(*)}$ mesons strong enough to form bound states but not strong enough in the isovector case. A recent analysis on lattice~\cite{Bicudo:2021qxj} shows that the meson-meson component in the $I(J^P)=0(1^+)$ $\bar b\bar b ud$ state has a fraction around 60\%.

\subsubsection{Compact tetraquark states}

Besides the molecular assignment, many works have investigated the compact doubly heavy tetraquarks via various methods, including quark potential models~\cite{Ader:1981db,Ballot:1983iv,Zouzou:1986qh,Heller:1986bt,Brink:1998as,Silvestre-Brac:1993wyf,Silvestre-Brac:1993zem,Semay:1994ht,Schaffner-Bielich:1998ogl,Czarnecki:2017vco,Vijande:2009kj,Hyodo:2012pm,Xing:2018bqt,Cui:2006mp,Silvestre-Brac:1992kaa,Hyodo:2017hue,Meng:2021yjr}, quark models with heavy quark symmetries~\cite{Gelman:2002wf,Eichten:2017ffp,Karliner:2017qjm}, QCD sum rules~\cite{Navarra:2007yw,Wang:2010uf,Dias:2011mi,Chen:2013aba,Wang:2017uld,Wang:2017dtg,Wang:2020jgb,Agaev:2018khe,Tang:2019nwv,Agaev:2019kkz,Agaev:2019lwh,Agaev:2020zag,Wang:2020jgb} and lattice QCD~\cite{Green:1998nt,Francis:2016hui,Cheung:2017tnt,Junnarkar:2018twb,Leskovec:2019ioa,Hudspith:2020tdf,Francis:2018jyb,Mohanta:2020eed}.

In the quark potential models, it was found that the $\bar Q\bar Q qq$ is possible to be bound below the two-meson threshold for certain potentials~\cite{Ader:1981db,Ballot:1983iv,Zouzou:1986qh,Heller:1986bt} but no bound states can be found if all quarks have the same mass, which was confirmed recently in a four-body calculation~\cite{Czarnecki:2017vco}.  It was also found~\cite{Vijande:2009kj} that although the possible configurations of tetraquarks  proliferate, only five of the candidates are stable, namely, $cc\bar q\bar q$ with $J^P(L,S,I)=1^+(0,1,0)$ and $bb\bar q\bar q$ with $J^P(L,S,I)=1^+(0,1,0),\ 3^-(1,2,1),\ 0^+(0,0,0)$ and $1^-(1,0,0)$, where $L$ and $S$ refer to the orbital angular momentum and total spin of the two mesons that couple to the tetraquark, among which the last one has a molecular nature. 
The doubly charmed system was also confirmed to be bound in Ref.~\cite{Valcarce:2010zs} using different potential models while in Ref.~\cite{Pepin:1996id} only one doubly-bottomed system with $I(J^P)=0(1^+)$ was found to be located below the threshold of the corresponding meson pair. 
With an adequate treatment of the four-body dynamics in the quark model picture of tetraquark states, it was found in Ref.~\cite{Richard:2018yrm} that the $I(J^P)=0(1^+)$ $cc\bar u\bar d$ system is at the edge of binding while the doubly-bottomed system is easier to be bound. On the other hand, the analysis within the chiral SU(3) quark model~\cite{Zhang:2007mu} or the relativistic quark model~\cite{Ebert:2007rn} found that the $I(J^P)=0(1^+)$ $cc\bar u\bar d$ state is not bound but above the thresholds for decays into open charm mesons. In Ref.~\cite{Meng:2020knc}, two bound states of $I(J^P)=0(1^+)$ $bb\bar u\bar d$ were found in a constituent quark model, one deeply bound compact tetraquark and one $BB^*$ molecule.

In Ref.~\cite{Lipkin:1986dw}, the masses of tetraquark states were obtained model-independently with known hadron masses as input and it was found that four-quark states containing two identical heavy quarks have a good probability of being stable against strong decay. In Ref.~\cite{Karliner:2017qjm}, the ground states of hadrons are described by the quark model and therein the mass of the predicted $J^P=1^+$ $cc\bar u\bar d$ state reads $(3882\pm12)$~MeV using the measured $\Xi_{cc}^{++}$ mass~\cite{LHCb:2017iph} as input, which nicely covers the LHCb result. 
Ref.~\cite{Karliner:2017qjm} also predicts a $1^+$ $bb\bar u\bar d$ state to be at $(10389\pm12)$~MeV, which is well below the $BB^*$ threshold and thus stable under strong and electromagnetic interactions.

By implementing the heavy antiquark-diquark symmetry~\cite{Savage:1990di}, the mass of doubly-heavy tetraquark states may be predicted using the relation
\begin{align}
    m(QQ\bar q\bar q)-m(QQq) \approx m(Qqq)-m(Q\bar q),
\end{align}
the essence of which can be traced back to Refs.~\cite{Eichten:1987xu,Lepage:1987gg}. For instance, after the double-charm baryon $\Xi_{cc}^{++}$ was observed~\cite{LHCb:2017iph}, the lightest double-bottom tetraquark state was predicted in Ref.~\cite{Eichten:2017ffp} to be at 10482~MeV, which is below the $BB^*$ threshold and stable under strong and electromagnetic interactions, with $J^P=1^+$, using the doubly-bottom baryon masses in Ref.~\cite{Karliner:2014gca} as input. The mass of $cc\bar u\bar d$ was determined in this way to be 3845 or 3905~MeV~\cite{Gelman:2002wf}, 3978~MeV~\cite{Eichten:2017ffp}, $(3947\pm11)$~MeV~\cite{Braaten:2020nwp}, 3900~MeV~\cite{Karliner:2013dqa} and 3929~MeV~\cite{Cheng:2020wxa}, see also the discussion in Ref.~\cite{Mehen:2017nrh}. 
The spectrum of some other doubly-heavy tetraquark states were also obtained in these works, including stable $bb\bar q\bar q$ tetraquark states.
Note that different from the case of doubly-heavy baryons, where the heavy diquark must be in a color anti-triplet, the two heavy quarks inside a doubly-heavy tetraquark can be in either a color anti-triplet or a color sextet. The approximate heavy antiquark-diquark symmetry is applicable only when the color sextet component of the diquark can be neglected and when the two heavy quarks are close to each other so that they acts as a pointlike color-antitriplet source just like a heavy antiquark. Thus, it cannot be applied to relate the heavy-heavy molecular systems to singly-heavy baryons.
Tetraquarks with the two charm quarks as a compact diquark may exist in addition to the molecular states. The mixing of these two configurations will make the spectrum more complicated.

The spectrum of $QQ\bar q\bar q$ tetraquark states are explored via the method of QCD sum rules in Refs.~\cite{Navarra:2007yw,Wang:2010uf,Dias:2011mi,Chen:2013aba,Wang:2017uld,Wang:2017dtg,Wang:2020jgb,Agaev:2018khe,Tang:2019nwv,Agaev:2019kkz,Agaev:2019lwh,Agaev:2020zag,Wang:2020jgb}. In Ref.~\cite{Dias:2011mi} it was argued that the molecular current of $DD^*$ yields a similar mass with that of the $D\bar D^*+c.c.$ molecule and thus the result will perfectly match $T_{cc}^+$ mass measured by LHCb if $X(3872)$ is a $D\bar D^*+c.c.$ molecule. 
The mass of the $cc\bar u\bar d$ ground state with $J^P=1^+$ was determined to be $(3.90\pm0.09)$~GeV in Refs.~\cite{Wang:2017uld,Wang:2017dtg}, which is also consistent with LHCb result. While in other works, it was found that usually the $J^P=1^+$ $bb\bar u\bar d$ state lies below the threshold of $BB^*$~\cite{Navarra:2007yw,Agaev:2018khe,Tang:2019nwv} while the charmed one is above the corresponding $DD^*$ threshold~\cite{Navarra:2007yw,Wang:2010uf,Tang:2019nwv}.

The explorations of doubly-heavy compact tetraquark states on lattice can be seen in, e.g., Refs.~\cite{Green:1998nt,Francis:2016hui,Cheung:2017tnt,Junnarkar:2018twb,Francis:2018jyb,Leskovec:2019ioa,Hudspith:2020tdf,Mohanta:2020eed,Bicudo:2017szl}. Deeply bound tetraquark states, $ud\bar b\bar b$, $qs\bar b\bar b$ and $ud\bar b\bar c$ with $J^P=1^+$ are predicted in Refs.~\cite{Francis:2016hui,Francis:2018jyb}, which are $189\pm10$, $98\pm7$  and $15\sim61$~MeV below the corresponding free two-meson thresholds, respectively. Ref.~\cite{Junnarkar:2018twb} obtained similar results for the above systems and predicted another three bound states of $uc\bar b\bar b$, $ud\bar c\bar c$ and $us\bar c\bar c$ located just below the corresponding free two-meson thresholds, and the masses of the double-charm states are in agreement with those obtained in Ref.~\cite{Cheung:2017tnt}. In Ref.~\cite{Mohanta:2020eed}, the tetraquark operators for $bb\bar u\bar d$ are constructed in both diquark-antidiquark ($[bb][\bar u\bar d]$) and molecular ($[b\bar u][b\bar d]$) configurations and two states were found on lattice, a compact tetraquark located at $189\pm18$~MeV below $BB^*$ threshold and a molecular one $17\pm14$~MeV above the same threshold.

\subsection{Heavy meson-baryon molecules v.s. doubly heavy baryons}

Different from the meson-meson case, from Table~\ref{tab:potentialsc} we can see that double-charm meson-baryon systems are more attractive or less repulsive than the hidden-charm ones. Specifically, the $D^{(*)}\Sigma_c^{(*)}$ systems with isospin-$1/2$ are more attractive than the isospin-$1/2$ $\bar D^{(*)}\Sigma_c^{(*)}$ systems while the latter are widely believed to be able to form bound states with experimental candidates, namely, the famous $P_c$ states~\cite{Aaij:2019vzc}. Therefore, it is natural that more deeply bound states of $D^{(*)}\Sigma_c^{(*)}$ exist, see Fig.~\ref{fig:specDBm5}. Similar conclusions can be drawn for $D_{1,2}\Sigma_c^{(*)}$ systems, as shown in Fig.~\ref{fig:specDBp5}, since they have the same form of leading order interactions. Other systems including $D^{(*)}\Xi_c^{(\prime*)}$, $D_{1,2}\Xi_c^{(\prime*)}$, $D_s^{(*)}\Omega_c^{(*)}$ and $D_{s1,s2}\Omega_c^{(*)}$ are also predicted to be bound easily, the spectra of which are shown in Figs.~\ref{fig:specDBm0} and \ref{fig:specDBp0}.

Similarly with the coupled channel analysis used in the pioneering works~\cite{Wu:2010jy,Wu:2010vk} of pentaquark states where the $P_c$ states were successfully predicted, Ref.~\cite{Dias:2018qhp} extended such a study to the double-charm systems and some deeply bound states of $D^{(*)}\Sigma_c^{(*)}$ with binding energies of $\mathcal{O}$(100 MeV) were found. It was extended further to the charm-bottom system and double-bottom systems in Refs.~\cite{Yu:2019yfr,Dias:2019klk}, and more deeply bound states of $D^{(*)}\Sigma_b^{(*)}$ and $\bar B^{(*)}\Sigma_c^{(*)}$ with binding energies of $\mathcal{O}$(300~MeV) and $\bar B^{(*)}\Sigma_b^{(*)}$  with binding energies of $\mathcal{O}$(400~MeV) were obtained. There it was also found that there are poles located about 100~MeV below the $\Lambda_b D$, $\Lambda_c\bar B$ and $\Lambda_b\bar B$ thresholds, respectively. Such conclusions are qualitatively consistent with our results that $D^{(*)}\Sigma_c^{(*)}$ and $D^{(*)}\Xi_c^{(*)}$ are attractive and the former is stronger. The meson-baryon transitions between the coupled channels $J / \psi N$-$\Lambda_{c} \bar{D}^{(*)}$-$\Sigma_{c}^{(*)} \bar{D}^{(*)}$ were explored in Ref.~\cite{Shimizu:2017xrg} and it was found that a doubly-charmed state, $\Xi_{cc}^*(4380)$, exists with almost the same mass as $P_c(4380)$. An S-wave scattering of ground state doubly-charmed baryons with the light pseudoscalar mesons were first investigated in Ref.~\cite{Guo:2011dd} and then in Ref.~\cite{Guo:2017vcf} by means of unitarized chiral effective theory and several doubly charmed baryon resonances were predicted. The spectrum was modified by including the effects of the $P$-wave excitation inside the charm diquark in Ref.~\cite{Yan:2018zdt}.

The interaction between $D\Lambda_c$ or $\bar B\Lambda_b$ from two-pion exchange are investigated in Ref.~\cite{Xu:2010fc} and it was claimed that a $\bar B\Lambda_b$ bound state from such an interaction is possible. In Ref.~\cite{Chen:2017vai}, the $D\Lambda_{c/b}$, $\bar B\Lambda_{c/b}$ systems were found possible to be bound by the $\sigma/\omega$ exchange interaction. Systematic studies on the $\Sigma_{c}^{(*)} D^{(*)}$ interactions within the framework of chiral effective field theory~\cite{Chen:2021htr} or one boson exchange~\cite{Liu:2020nil} were performed, and the $I=1/2$ systems may form bound states with binding energies of about several or dozens MeV, consistent with the results obtained here, see Table~\ref{tab:pole05m}. Therein, deeper bound states of the $\Sigma_{c}^{(*)} \bar{B}^{(*)}, \Sigma_{b}^{(*)} D^{(*)}$ and $\Sigma_{b}^{(*)} \bar{B}^{(*)}$ systems were also predicted to exist due to the larger reduced masses, as expected.

The doubly-heavy pentaquark states with compact configurations were explored within the color-magnetic interaction model~\cite{Zhou:2018bkn}, non-relativistic
constituent quark model~\cite{Zhu:2019iwm} and QCD sum rules~\cite{Xing:2021yid,Wang:2018lhz} where some narrow exotic pentaquark states were predicted, constituent quark model~\cite{Park:2018oib} where one strong-interaction stable state, $ccud\bar s$ was predicted.

From Table~\ref{tab:pole05m} and Fig.~\ref{fig:specDBm5}, one sees that the lightest double-charm meson-baryon molecule is the one in the $D\Lambda_c$ system. The state has a mass large enough for it to decay into $\Xi_{cc}\pi$ and could be broad. So are the other similar states.

\subsection{Heavy di-baryons}

For heavy di-baryons, the leading order interaction from the vector meson exchange leads to evident binding only for the $\Sigma_c^{(*)}\Sigma_c^{(*)}$ systems, see Table~\ref{tab:pole0p} and Fig.~\ref{fig:specBB0}. The attractions for the $\Xi_c\Xi_c$ related systems are too weak and only remote virtual poles are found, which are not robust somehow and can get sizeably modified by the omitted momentum-dependent terms. 

In our simple model the leading order interaction for the $\Lambda_c\Lambda_c$ system is repulsive and thus it cannot be bound. Within various models, it was found that the $\Lambda_c\Lambda_c$ system can not be bound by itself~\cite{Lee:2011rka,Meguro:2011nr,Huang:2013rla,Oka:2013xxa,Carames:2015sya,Garcilazo:2020acl} but Refs.~\cite{Meguro:2011nr,Huang:2013rla,Oka:2013xxa,Li:2012bt} showed that the coupling to the strongly attractive $\Sigma_c^{(*)}\Sigma_c^{(*)}$ system may lead to a states below $\Lambda_c\Lambda_c$ threshold . In an analysis with QCD sum rules~\cite{Wang:2021qmn}, the states with the same quark components as $\Lambda_c\Lambda_c$ were found all above the $\Lambda_c\Lambda_c$ threshold. In Refs.~\cite{Gerasyuta:2011zx,Lu:2017dvm,Chen:2017vai}, on the contrary, the single channel $\Lambda_c\Lambda_c$ can be bound by itself; however, Ref.~\cite{Lu:2017dvm} warns that a more thorough theoretical exploration is needed to determine whether the $\Lambda_c\Lambda_c$ system really binds.

Some other doubly-heavy di-baryon systems were also explored. Several realistic phenomenological nucleon-nucleon interaction models are employed in Ref.~\cite{Froemel:2004ea}. It was found there that the $\Xi^{(\prime)}_c\Xi^{(\prime)}_c$ and $\Sigma_c\Sigma_c$ systems can be bound in some models. 
From one-boson exchange with coupled channel effects included, Refs.~\cite{Lee:2011rka,Yang:2018amd} found that $\Xi_c^{(\prime*)}\Xi_c^{(\prime*)}$ and $\Omega_c^{(*)}\Omega_c^{(*)}$ may be loosely bound while the isoscalar $\Sigma_c^{(*)}\Sigma_c^{(*)}$ systems can be deeply bound. Similar conclusion was obtained in Ref.~\cite{Huang:2013rla} within the framework of quark delocalization color screening model that the $\Sigma_c^{(*)}\Sigma_c^{(*)}$ single-channel system can be deeply bound.  
In Ref.~\cite{Garcilazo:2020acl}, the potential of  $\Sigma_c\Sigma_c$ was derived from a constituent quark model and a bound state with $I(J^P)=0(0^+)$ was obtained with a binding energy about 6.2~MeV. The  long-range pion exchange force is strong enough to form molecules of $\left[\Sigma_{Q} \Xi_{Q}^{\prime}\right]_{J=1}^{I=1 / 2}$, $\left[\Sigma_{Q} \Lambda_{Q}\right]_{J=1}^{I=1}(Q=b, c)$, $\left[\Sigma_{b} \Xi_{b}^{\prime}\right]_{J=1}^{I=3 / 2}$ and $\left[\Xi_{b} \Xi_{b}^{\prime}\right]_{J=1}^{I=0}$ where the $S$-$D$ mixing is necessary~\cite{Li:2012bt}. In Ref.~\cite{Lu:2017dvm}, the $\Sigma_{Q}^{(*)} \Sigma_{Q}^{(*)}$ were claimed to be good candidates of bound states from the one-pion and vector meson exchanges while it concludes that a more thorough analysis is necessary to determine whether there is a binding for the $\Lambda_{Q} \Sigma_{Q}^{(*)}$.

The di-baryon systems with two heavy quarks were investigated in Ref.~\cite{Vijande:2016nzk} with a simple quark model but no bound or metastable state was found.

\section{Summary}\label{sec:5}

In this work we have obtained an overall spectrum of hadronic molecules composed of a pair of charmed hadrons, including all the $S$-wave singly-charmed mesons and baryons as well as the $s_\ell=1/2$ $P$-wave charmed mesons. The interaction is assumed to be dominated by the light vector meson exchange and approximated by constants at leading order, which are derived systematically from the couplings that satisfy HQSS
and SU(3) flavor symmetry.

One should keep in mind that the spectrum predicted here should be regarded as the leading approximation of the spectrum for heavy-heavy molecular states, which gives only a general overall feature of the heavy-heavy hadronic molecular spectrum. The numerical results can receive large quantitative corrections { due to the limitations of our treatments, which we discuss qualitatively in the following.

\begin{itemize}

    \item We have only considered the leading interactions described by constant contact terms. The momentum dependent terms (including both spin-dependent and spin-independent contributions) may change the spectrum we obtained visibly, especially for the systems where the poles are far away from the corresponding thresholds. The spin-dependent terms will also lift the degeneracy of the same system with different total spins.
    \item The coupled channel effects have been neglected. In some cases the coupled-channel effects may play an important role in the formation of near threshold states. However, it is common and natural that the near-threshold pole found in a coupled-channel system dominantly couples to a single channel, see, e.g., the $D_{s0}^*(2317)$, which is dynamically generated in the $DK$ and $D_s\eta$ system but couples dominantly to $DK$~\cite{Kolomeitsev:2003ac,Guo:2006fu,Gamermann:2006nm}, and the $\Xi_{cc}(4083)$ state with $J^P=1/2^-$, which is dynamically generated in the $\Sigma_c D$ and $\Xi_c'D_s$ system but couples dominantly to $\Sigma_c D$~\cite{Dias:2018qhp}.\footnote{ The $\Sigma_c D$ bound state obtained in this work has a mass around 4.3~GeV, much closer to the threshold; see Table~\ref{tab:pole05m} and Fig.~\ref{fig:specDBm5}.}
    \item The hadronic molecules shown in Figs.~\ref{fig:specDBm5}-\ref{fig:specDBp0} can couple to normal double-charm baryons as well as channels with a double-charm baryon and a light meson. It is expected that each of these two types of systems also forms a spectrum. The physical spectrum of double-charm baryons should incorporates the mixing among all the three spectra. Coupled channels including both the charm-baryon--charm-meson channels and light-meson--double-charm-baryon channels have been considered in, e.g., Ref.~\cite{Dias:2018qhp} for the $\Xi_{cc}$ type molecular states. Mixing of light-meson--double-charm-baryon molecular states with the normal double-charm baryons with a $P$-wave excitation inside the charm diquark has been considered in Ref.~\cite{Yan:2018zdt}. Yet, a model considering all the three kinds of channels does not exist so far.
    \item The exchange of other particles, such as the light scalar mesons, charmed mesons and charmonia, are not considered, the effect of which can be partly covered by varying the cutoff. In addition, the interactions considered here are of leading order in the $1/N_c$ expansion, where $N_c$ is the number of colors, i.e., the Okubo-Zweig-Iizuka violating interactions have been neglected. Such contributions will also lift the degeneracy of the same system with different total spins.
\end{itemize}
}
% since we have not considered the momentum-dependent terms (including both spin-dependent and spin-independent contributions) that are of higher order in the very near-threshold region, or the effects of coupled channels, which may be important in some systems, or mixing with other non-molecular components. 
% {\color{blue}Such contributions will also lift the degeneracy of the same system belonging to different spin multiples.} 
Therefore, {although we expect the spectrum given here should present an overall pattern of the hadronic molecules formed by a pair of charmed mesons and/or baryons}, specific systems may {quantitatively} differ from the predictions here due to the limitations of our treatments.

In total we obtained 124 double-charm hadronic molecules and we summarize the main feature of this spectrum in the following.
\begin{enumerate}
    \item Unlike the isovector $D^*D^{(*)}$ systems that are repulsive, the isoscalar ones have attractive interaction from the light vector meson exchange and the total potential makes the systems at the edge of forming near-threshold molecules. With a reasonable cutoff regularizing the loop integral, the binding energy of the $I(J^P)=0(1^+)$ $DD^*$ system is consistent with the double charm tetraquark $T_{cc}^+$ in LHCb observation. If the hadronic molecule structure of the $T_{cc}^+$ is confirmed, which is rather natural given the closeness to the $DD^*$ threshold, many other similar states with $I=0$ including $D^*D^*$, $D^{(*)}D_{1,2}$, $D_{1,2}D_{1,2}$ can also exist.
    \item Given that the famous $P_c$ states are hadronic molecules of $\bar D^{(*)}\Sigma_c^{(*)}$, it is natural to expect the existence of double-charm $D^{(*)}\Sigma_c^{(*)}$ and $D_{1,2}\Sigma_c^{(*)}$ states with $I=1/2$ because the attraction from the light vector meson exchange for the latter is stronger than that for the former. Similar conclusions can be made for the $D^{(*)}\Xi_c^{(\prime*)}$ and $D_{1,2}\Xi_c^{(\prime*)}$ channels, especially when the $P_{cs}$ states are established experimentally. The hadronic molecules in other systems including $D^{(*)}\Lambda_c$, $D_{1,2}\Lambda_c$, $D_s^{(*)}\Xi_c^{(\prime*)}$, $D_{s1,2}\Xi_c^{(\prime*)}$ $D_s^{(*)}\Omega_c^{(*)}$ and $D_{s1,2}\Omega_c^{(*)}$ are also predicted.
    \item Within our simple model, in the double-charm di-baryon sector, only isoscalar $\Sigma_c^{(*)}\Sigma_c^{(*)}$ systems are expected to be good candidates of bound di-baryon states. As discussed in the literature, the inclusion of other contributions to the interaction as well as the coupling to other channels may make additional di-baryon bound states possible.
\end{enumerate}

Due to the heavy quark flavor symmetry, the potentials in the bottom sector are the same
as those in the charm sector, if using the nonrelativistic field normalization, and we expect at least the same number of molecular states in the analogous systems therein. Because of the much heavier reduced masses of hidden-bottom system, it will be easier to form bound states than the charmed systems, and there may even be excited states if the ground states are deeply bound, the treatment of which, however, requires momentum-dependent interactions and is beyond the scope of this paper.

\begin{acknowledgments}
We would like to thank Marek Karliner {and Eulogio Oset} for useful comments and Ming-Zhu Liu for pointing out a mistake about the quantum numbers of a pair of identical fermions.
This work is supported in part by the Chinese Academy of Sciences (CAS) under Grant No.~XDPB15, No.~XDB34030000 and No.~QYZDB-SSW-SYS013, by the National Natural Science Foundation of China (NSFC) under Grant No.~11835015, No.~12047503 and No.~11961141012, and by the NSFC and the Deutsche Forschungsgemeinschaft (DFG, German Research Foundation) through funds provided to the Sino-German Collaborative Research Center ``Symmetries and the Emergence of Structure in QCD'' (NSFC Grant No. 12070131001, DFG Project-ID 196253076 -- TRR110).
\end{acknowledgments}

%---------------------------------------charmed---------------------------------------------
\appendix
\section{The flavor factor $F$}\label{app:poten}

In this section we list the factor $F$ that accounts for the flavor information for the exchange of different mesons in different systems. The essential point is that the vector mesons, $\rho,\omega$ and $\phi$, only couple to the light quark in heavy or antiheavy hadron. The conclusion is that the isoscalar vector, $\omega$ and $\phi$, exchange have opposite signs for $qq$ and $q\bar q$ interactions while the isovector meson, $\rho$, exchange has the same sign in these two cases. Note that the negative $C$-parity of vector mesons has been taken into account. The conclusion for isoscalar meson exchange is apparent and in the following we give a brief deduction for the case of isovector exchange.

For simplicity, we consider the hadrons that contain only one $u$ or $d$ quark. The Lagrangian for the isospin structure, taking the $\rho$ meson exchange for example, reads
\begin{equation}
    \mathcal L=\psi^\dagger{\bm {\rho\cdot\tau}}\psi+\rm c.c.
\end{equation}
with $\psi=(u,d)^T$, $\psi^\dagger=(u^\dagger,d^\dagger)$, $\bm \tau$ the Pauli matrices, where $T$ means transpose. Note that here we are only interested in the flavor structure and hence we denote the field creating a $u$ quark by $u^\dagger$ instead of $\bar u$ to avoid confusion with the field for antiquark below. The charge conjugation term, which needs special attentions, is
\begin{equation}
    c\bar\psi^{'\dagger}(\bm {\rho\cdot\tau})^T\bar\psi'
\end{equation}
with $\bar\psi'=(\bar u,\bar d)^T$, $\bar\psi^{'\dagger}=(\bar u^\dagger,\bar d^\dagger)$ and $c=-1$ the charge conjugation factor for the $\rho$ meson. Using the common convention for the isospin eigenstates,
\begin{equation}
    \ket{u}=\ket{\uparrow},
    \ket{d}=\ket{\downarrow},
    \ket{\bar d}=\ket{\uparrow},
    \ket{\bar u}=-\ket{\downarrow},
\end{equation}
we rearrange the charge conjugation term as 
\begin{equation}
    c\bar\psi^{\dagger}\tau_2(\bm {\rho\cdot\tau})^T\tau_2\bar\psi=-c\bar\psi^{\dagger}(\bm {\rho\cdot\tau})\bar\psi
\end{equation}
where $\bar\psi=(\bar d, -\bar u)^T=i\tau_2\bar\psi'$. Now we are ready to calculate the isospin factors for $qq$ and $q\bar q$ interactions via the isovector meson exchange. Explicitly, they are given by
\begin{align}
    V_{qq}&\propto\bm{\tau_1\cdot\tau_2}=\left\{\begin{array}{ll}
        1& \text{for }I=1\\
        -3& \text{for }I=0
    \end{array}\right. ,\\
    V_{q\bar q}&=-cV_{qq}.
\end{align}

\begin{table*}[tbhp]
\caption{The group theory factor $F$, defined in Eq.~\eqref{eq:potential}, for the interaction of charm-anticharm/charm-charm hadron pairs with only the light vector-meson exchanges. Here both charm hadrons are the $S$-wave ground states. $I$ is the isospin and $S$ is the strangeness. { Note that we have collect the pairs with the heavy hadrons in the same spin multiples (such as $DD$, $DD^*$, etc.) in one row, and the several numbers in the column of ``Thresholds" represent the thresholds of these different pairs in an increasing order.} Positive $F$ means that the interaction attractive. {The values in the column of ``$F$" correspond to those for the exchanged particles in the column of ``Exchanged particles" in order.}} \label{tab:potentialsc}
\centering
\begin{ruledtabular}
\begin{tabular}{l|ccccc}
System &  $I$&$S$ & Thresholds [MeV] & Exchanged particles & $F$\\
\hline
$D^{(*)}\bar D^{(*)}/D^{(*)} D^{(*)}$&1&0/0&$(3734,3876,4017)$ &$\rho,\omega$ & $-\frac12,\frac12/-\frac12,-\frac12$\\
&  0&& & &$\frac32,\frac12$/$\frac32,-\frac12$ \\
$D_s^{(*)}\bar D^{(*)}$/$D_s^{(*)} D^{(*)}$& $\frac12$&1/1&$(3836,3977,3979,4121)$&${K^*}$ & $0${/$-1$}\\
$D^{(*)}_s\bar D^{(*)}_s $/$D^{(*)}_s D^{(*)}_s $& 0&0/2&$(3937,4081,4224)$&$\phi$ & $1$/$-1$\\
\hline
$\bar D^{(*)}\Lambda_c$/$ D^{(*)}\Lambda_c$& $\frac12$&0/0 &$(4154,4295)$&$\omega$ & $-1$/$1$\\
$\bar D_s^{(*)}\Lambda_c$/$ D_s^{(*)}\Lambda_c$&0& $-1/1$ &$(4255,4399)$&$-$ & $0$/$0$\\
$\bar D^{(*)}\Xi_c$/$ D^{(*)}\Xi_c$&1& $-1/-1$ &$(4337,4478)$ &$\rho,\omega$ & $-\frac12,-\frac12$/$-\frac12,\frac12$\\
& 0& & & & $\frac32,-\frac12/\frac32,\frac12$\\
$\bar D_s^{(*)}\Xi_c$/$ D_s^{(*)}\Xi_c$& $\frac12$&$-2/0$& $(4438,4582)$&$\phi$ & $-1$/$1$\\
\hline
$\bar D^{(*)}\Sigma_c^{(*)}$/$ D^{(*)}\Sigma_c^{(*)}$& $\frac32$&0/0&$(4321,4385,4462,4527)$ &$\rho,\omega$ & $-1,-1$/$-1,1$\\
& $\frac12$&& & & $2,-1$/2,1\\
$\bar D_s^{(*)}\Sigma_c^{(*)}$/$D_s^{(*)}\Sigma_c^{(*)}$&1& $-1/1$&$(4422,4486,4566,4630)$ &$-$ & $0/0$\\
$\bar D^{(*)}\Xi_c^{'(*)}$/$ D^{(*)}\Xi_c^{'(*)}$&1& $-1/-1$&$(4446,4513,4587,4655)$ &$\rho,\omega$ & $-\frac12,-\frac12$/$-\frac12,\frac12$\\
& 0&& && $\frac32,-\frac12$/$\frac32,\frac12$\\
$\bar D_s^{(*)}\Xi_c^{'(*)}$/$D_s^{(*)}\Xi_c^{'(*)}$& $\frac12$&$-2/0$& $(4547,4614,4691,4758)$&$\phi$ & $-1/1$\\
$\bar D^{(*)}\Omega_c^{(*)}$/$D^{(*)}\Omega_c^{(*)}$& $\frac12$&$-2/0$&$(4562,4633,4704,4774)$ &$-$ & $0/0$\\
$\bar D_s^{(*)}\Omega_c^{(*)}$/$D_s^{(*)}\Omega_c^{(*)}$&0& $-3/-1$&$(4664,4734,4807,4878)$ &$\phi$ & $-2/2$\\

\hline
$ \Lambda_c\bar\Lambda_c$/$ \Lambda_c\Lambda_c$& 0&0/0&$(4573)$ &$\omega$ & $2$/$-2$\\
$\Lambda_c\bar \Xi_c$/$\Lambda_c \Xi_c$& $\frac12$&$1/-1$&$(4756)$ &$\omega{/K^*}$ & $1{,0}$/$-1{,-1}$\\
$\Xi_c\bar \Xi_c$/$\Xi_c \Xi_c$&1& $0/-2$&$(4939)$ &$\rho,\omega,\phi$ & $-\frac12,\frac12,1$/$-\frac12,-\frac12,-1$\\
& 0& & & & $\frac32,\frac12,1$/$\frac32,-\frac12,-1$\\
\hline
$\Lambda_c\bar\Sigma_c^{(*)}$/$\Lambda_c\Sigma_c^{(*)}$& 1&0/0&$(4740,4805)$ &$\omega{/K^*}$ & $1{,0}$/$-1{,-1}$\\

$\Lambda_c\bar\Xi_c^{'(*)}$/$\Lambda_c\Xi_c^{'(*)}$&$\frac12$&$1/-1$&$(4865,4932)$ &$\omega$ & $1$/$-1$\\

$\Lambda_c\bar\Omega_c^{(*)}$/$\Lambda_c\Omega_c^{(*)}$ &0&$2/-2$&$(4982,5052)$ &$-$ & $0$/$0$\\

$\Sigma_c^{(*)}\bar\Xi_c $/$\Sigma_c^{(*)}\Xi_c $  &$\frac32$&$1/-1$&$(4923,4988)$ &$\rho,\omega,K^*$ & $-1,1,0/-1,-1,-2$\\

&$\frac12$&& && $2,1,0$/$2,-1,-2$\\
$\Xi_c \bar\Xi_c^{'(*)}$/$\Xi_c \Xi_c^{'(*)}$ &1&$0/-2$&$(5048,5115)$ &$\rho,\omega,\phi$ & $-\frac12,\frac12,1$/$-\frac12,-\frac12,-1$\\

& 0&& & & $\frac32,\frac12,1$/$\frac32,-\frac12,-1$\\
$\Xi_c \bar\Omega_c^{(*)}$/$\Xi_c \Omega_c^{(*)}$ &$\frac12$&$1/-3$&$(5165,5235)$ &$\phi,K^*$ & $2,0$/$-2,-2$\\
\hline
$\Sigma_c^{(*)}\bar\Sigma_c^{(*)}$/$\Sigma_c^{(*)}\Sigma_c^{(*)}$  & 2&0/0&$(4907,4972,5036)$ &$\rho,\omega$ & $-2,2$/$-2,-2$\\

 & 1&& & & $2,2$/$2,-2$\\

 & 0&& & & $4,2$/$4,-2$\\

$\Sigma_c^{(*)}\bar\Xi^{'(*)}_c$/$\Sigma_c^{(*)}\Xi^{'(*)}_c$  &$\frac32$&$1/-1$&$(5032,5097,5100,5164)$ &$\rho,\omega,K^*$ & $-1,1,0$/$-1,-1-2$\\

 & $\frac12$&& & & $2,1,0$/$2,-1,-2$\\

$\Sigma_c^{(*)}\bar\Omega^{(*)}_c$/$\Sigma_c^{(*)}\Omega^{(*)}_c$  &0&$2/-2$& $(5149,5213,5219,5284)$&$-$ & $0$/$0$\\

$\Xi_c^{'(*)} \bar\Xi_c^{'(*)}$/$\Xi_c^{'(*)}\Xi_c^{'(*)}$&1&$0/-2$&$(5158,5225,5292)$ &$\rho,\omega,\phi$ & $-\frac12,\frac12,1$/$-\frac12,-\frac12,-1$\\

 &0&& & & $\frac32,\frac12,1$/$\frac32,-\frac12,-1$\\

$\Xi^{'(*)}_c \bar\Omega_c^{(*)}$/$\Xi^{'(*)}_c\Omega_c^{(*)}$&$\frac12$&$1/-3$&$(5272,5341,5345,5412)$ &$\phi,K^*$ & $2,0$/$-2,-2$\\

$\Omega_c ^{(*)}\bar\Omega_c^{(*)}$/$\Omega_c ^{(*)}\Omega_c^{(*)}$  &0&$0/-4$&$(5390,5461,5532)$ &$\phi$ & $4$/$-4$
\end{tabular}
\end{ruledtabular}
\end{table*}

\begin{table*}
\caption{The group theory factor $F$, defined in Eq.~\eqref{eq:potential}, for the interaction of charm-anticharm/charm-charm  hadron pairs with only the light vector-meson exchanges. Here one of the charm hadrons is an $s_\ell=3/2$ charm meson. See the caption of Table~\ref{tab:potentialsc}. } \label{tab:potentialsc1}
\centering
\begin{ruledtabular}
\begin{tabular}{l|ccccc}
System &  $I$&$S$ & Thresholds [MeV] & Exchanged particles & $F$ \\
\hline
$D^{(*)}\bar D_{1,2}$/$D^{(*)} D_{1,2}$& 0&0/0&$(4289,4330,4431,4472)$ &$\rho,\omega$ & $\frac32,\frac12$/$\frac32,-\frac12$\\
& 1&0/0& & & $-\frac12,\frac12$/ $-\frac12,-\frac12$\\
$D^{(*)}\bar D_{s1,s2}$/$D^{(*)} D_{s1,s2}$& $\frac12$&$1/-1$&$(4390,4431,4534,4575)$&$-$ & $0$/$0$\\
$D_s^{(*)}\bar D_{1,2}$/$D_s^{(*)} D_{1,2}$& $\frac12$&$-1/1$&$(4402,4436,4544,4578)$&$-$ & $0$/$0$\\
$D^{(*)}_s\bar D_{s1,s2} $/$D^{(*)}_s D_{s1,s2} $&0& 0/$-2$&$(4503,4537,4647,4681)$&$\phi$ & $1$/$-1$\\
\hline
$D_{1,2}\bar D_{1,2}$/$D_{1,2} D_{1,2}$&0&0/0&$(4844,4885,4926)$ &$\rho,\omega$ & $\frac32,\frac12$/$\frac32,-\frac12$\\
& 1&& & & $-\frac12,\frac12$/$-\frac12,-\frac12$\\
$D_{s1,s2}\bar D_{1,2}$/$D_{s1,s2} D_{1,2}$& $\frac12$&1/1&$(4957,4991,4998,5032)$& & $0$/0\\
$D_{s1,s2}\bar D_{s1,s2} $/$D_{s1,s2} D_{s1,s2} $&0& $0/-2$&$(5070,5104,5138)$&$\phi$ & $1$/1\\
\hline
$\Lambda_c\bar D_{1,2}$/$\Lambda_c D_{1,2}$& $\frac12$&0/0 &$(4708,4750)$&$\omega$ & $-1$/1\\
$\Lambda_c\bar D_{s1,s2}$/$\Lambda_c D_{s1,s2}$&0& $-1/1$ &$(4822,4856)$&$-$ & $0$/0\\
$\Xi_c\bar D_{1,2}$/$\Xi_c D_{1,2}$& 1&$-1/-1$ &$(4891,4932)$ &$\rho,\omega$ & $-\frac12,-\frac12$/$-\frac12,\frac12$\\
& 0& & & & $\frac32,-\frac12$/$\frac32,\frac12$\\
$\Xi_c\bar D_{s1,s2}$/$\Xi_c D_{s1,s2}$& $\frac12$&$-2/0$& $(5005,5039)$&$\phi$ & $-1$/1\\
\hline
$\Sigma_c^{(*)}\bar D_{1,2}$/$\Sigma_c^{(*)} D_{1,2}$& $\frac32$&0/0&$(4876,4917,4940,4981)$ &$\rho,\omega$ & $-1,-1$/$-1,1$\\
& $\frac12$&& & & $2,-1$/2,1\\
$\Sigma_c^{(*)}\bar D_{s1,s2}$/$\Sigma_c^{(*)} D_{s1,s2}$&1& $1/-1$&$(4989,5023,5053,5087)$ &$-$ & $0$/0\\
$\Xi_c^{'(*)}\bar D_{1,2}$/$\Xi_c^{'(*)} D_{1,2}$&1& $-1/-1$&$(5001,5042,5068,5109)$ &$\rho,\omega$ & $-\frac12,-\frac12$/$-\frac12,\frac12$\\
& 0&& & & $\frac32,-\frac12$/$\frac32,\frac12$\\
$\Xi_c^{'(*)}\bar D_{s1,s2}$/$\Xi_c^{'(*)}D_{s1,s2}$& $\frac12$&$-2/0$& $(5114,5148,5181,5215)$&$\phi$ & $-1$/1\\
$\Omega_c^{(*)}\bar D_{1,2}$/$\Omega_c^{(*)} D_{1,2}$& $\frac12$&$-2/-2$&$(5117,5158,5188,5229)$ &$-$ & $0$/0\\
$\Omega_c^{(*)}\bar D_{s1,s2}$/$\Omega_c^{(*)} D_{s1,s2}$&0& $-3/-1$&$(5230,5264,5301,5335)$ &$\phi$ & $-2$/2\\
\end{tabular}
\end{ruledtabular}
\end{table*}
%-----------------------------------bottomed--------------------------------------
\begin{table*}[tbhp]
\caption{The group theory factor $F$, defined in Eq.~\eqref{eq:potential}, for the interaction of bottom-antibottom/bottom-bottom hadron pairs with only the light vector-meson exchanges. Here both bottom hadrons are the $S$-wave ground states. See the caption of Table~\ref{tab:potentialsc}. } \label{tab:potentialsb}
\centering
\begin{ruledtabular}
\begin{tabular}{l|ccccc}
System &  $I$&$S$ & Thresholds [MeV] & Exchanged particles & $F$\\
\hline
$B^{(*)}\bar B^{(*)}/B^{(*)} B^{(*)}$&1&0/0&$(10559, 10604, 10649)$ &$\rho,\omega$ & $-\frac12,\frac12/-\frac12,-\frac12$\\
&  0&& & &$\frac32,\frac12$/$\frac32,-\frac12$ \\
$B_s^{(*)}\bar B^{(*)}$/$B_s^{(*)} B^{(*)}$& $\frac12$&1/1&$(10646, 10695, 10692, 10740)$&${K^*}$ & $0$/${-1}$\\
$B^{(*)}_s\bar B^{(*)}_s $/$B^{(*)}_s B^{(*)}_s $& 0&0/2&$(10734, 10782, 10831)$&$\phi$ & $1$/$-1$\\
\hline
$\bar B^{(*)}\Lambda_b$/$ B^{(*)}\Lambda_b$& $\frac12$&0/0 &$(10899, 10944)$&$\omega$ & $-1$/$1$\\
$\bar B_s^{(*)}\Lambda_b$/$ B_s^{(*)}\Lambda_b$&0& $-1/1$ &$(10986, 11035)$&$-$ & $0$/$0$\\
$\bar B^{(*)}\Xi_b$/$ B^{(*)}\Xi_b$&1& $-1/-1$ &$(11074, 11119)$ &$\rho,\omega$ & $-\frac12,-\frac12$/$-\frac12,\frac12$\\
& 0& & & & $\frac32,-\frac12/\frac32,\frac12$\\
$\bar B_s^{(*)}\Xi_b$/$ B_s^{(*)}\Xi_b$& $\frac12$&$-2/0$& $(11161, 11210)$&$\phi$ & $-1$/$1$\\
\hline
$\bar B^{(*)}\Sigma_b^{(*)}$/$ B^{(*)}\Sigma_b^{(*)}$& $\frac32$&0/0&$(11093, 11138, 11112, 11157)$ &$\rho,\omega$ & $-1,-1$/$-1,1$\\
& $\frac12$&& & & $2,-1$/2,1\\
$\bar B_s^{(*)}\Sigma_b^{(*)}$/$B_s^{(*)}\Sigma_b^{(*)}$&1& $-1/1$&$(11180, 11228, 11199, 11248)$ &$-$ & $0/0$\\
$\bar B^{(*)}\Xi_b^{'(*)}$/$ B^{(*)}\Xi_b^{'(*)}$&1& $-1/-1$&$(11215, 11260, 11233, 11279)$ &$\rho,\omega$ & $-\frac12,-\frac12$/$-\frac12,\frac12$\\
& 0&& && $\frac32,-\frac12$/$\frac32,\frac12$\\
$\bar B_s^{(*)}\Xi_b^{'(*)}$/$B_s^{(*)}\Xi_b^{'(*)}$& $\frac12$&$-2/0$& $(11302, 11350, 11321, 11369)$&$\phi$ & $-1/1$\\
$\bar B^{(*)}\Omega_b^{(*)}$/$B^{(*)}\Omega_b^{(*)}$& $\frac12$&$-2/0$&$(11326, 11371, 11349, 11395)$ &$-$ & $0/0$\\
$\bar B_s^{(*)}\Omega_b^{(*)}$/$B_s^{(*)}\Omega_b^{(*)}$&0& $-3/-1$&$(11413, 11462, 11437, 11485)$ &$\phi$ & $-2/2$\\

\hline
$ \Lambda_b\bar\Lambda_b$/$ \Lambda_b\Lambda_b$& 0&0/0&$(11239)$ &$\omega$ & $2$/$-2$\\
$\Lambda_b\bar \Xi_b$/$\Lambda_b \Xi_b$& $\frac12$&$1/-1$&$(11414)$ &$\omega${, $K^*$} & $1{,0}$/$-1{,-1}$\\
$\Xi_b\bar \Xi_b$/$\Xi_b \Xi_b$&1& $0/-2$&$(11589)$ &$\rho,\omega,\phi$ & $-\frac12,\frac12,1$/$-\frac12,-\frac12,-1$\\
& 0& & & & $\frac32,\frac12,1$/$\frac32,-\frac12,-1$\\
\hline
$\Lambda_b\bar\Sigma_b^{(*)}$/$\Lambda_b\Sigma_b^{(*)}$& 1&0/0&$(11433, 11452)$ &$\omega$ & $2$/$-2$\\

$\Lambda_b\bar\Xi_b^{'(*)}$/$\Lambda_b\Xi_b^{'(*)}$&$\frac12$&$1/-1$&$(11555, 11573)$ &$\omega${, $K^*$} & $1{,0}$/$-1{,-1}$\\\\

$\Lambda_b\bar\Omega_b^{(*)}$/$\Lambda_b\Omega_b^{(*)}$ &0&$2/-2$&$(11666, 11690)$ &$-$ & $0$/$0$\\

$\Xi_b \bar\Sigma_b^{(*)}$/$\Xi_b \Sigma_b^{(*)}$  &$\frac32$&$-1/-1$&$(11608, 11627)$ &$\rho,\omega,K^*$ & $-1,1,0/-1,-1,-2$\\

&$\frac12$&& && $2,1,0$/$2,-1,-2$\\
$\Xi_b \bar\Xi_b^{'(*)}$/$\Xi_b \Xi_b^{'(*)}$ &1&$0/-2$&$(11729, 11748)$ &$\rho,\omega,\phi$ & $-\frac12,\frac12,1$/$-\frac12,-\frac12,-1$\\

& 0&& & & $\frac32,\frac12,1$/$\frac32,-\frac12,-1$\\
$\Xi_b \bar\Omega_b^{(*)}$/$\Xi_b \Omega_b^{(*)}$ &$\frac12$&$1/-3$&$(11841, 11864)$ &$\phi,K^*$ & $2,0$/$-2,-2$\\
\hline
$\Sigma_b^{(*)}\bar\Sigma_b^{(*)}$/$\Sigma_b^{(*)}\Sigma_b^{(*)}$  & 2&0/0&$(11626, 11646, 11665)$ &$\rho,\omega$ & $-2,2$/$-2,-2$\\

 & 1&& & & $2,2$/$2,-2$\\

 & 0&& & & $4,2$/$4,-2$\\

$\Sigma_b^{(*)}\bar\Xi^{'(*)}_b$/$\Sigma_b^{(*)}\Xi^{'(*)}_b$  &$\frac32$&$1/-1$&$(11748, 11768, 11767, 11786)$ &$\rho,\omega,K^*$ & $-1,1,0$/$-1,-1,-2$\\

 & $\frac12$&& & & $2,1,0$/$2,-1,-2$\\

$\Sigma_b^{(*)}\bar\Omega^{(*)}_b$/$\Sigma_b^{(*)}\Omega^{(*)}_b$  &0&$2/-2$& $(11859, 11879, 11883, 11903)$&$K^*$ & $0$/$-4$\\

$\Xi_b^{'(*)} \bar\Xi_b^{'(*)}$/$\Xi_b^{'(*)}\Xi_b^{'(*)}$&1&$0/-2$&$(11870, 11889, 11908)$ &$\rho,\omega,\phi$ & $-\frac12,\frac12,1$/$-\frac12,-\frac12,-1$\\

 &0&& & & $\frac32,\frac12,1$/$\frac32,-\frac12,-1$\\

$\Xi^{'(*)}_b \bar\Omega_b^{(*)}$/$\Xi^{'(*)}_b\Omega_b^{(*)}$&$\frac12$&$1/-3$&$(11981, 12000, 12005, 12024)$ &$\phi,K^*$ & $2,0$/$-2,-2$\\

$\Omega_b ^{(*)}\bar\Omega_b^{(*)}$/$\Omega_b ^{(*)}\Omega_b^{(*)}$  &0&$0/-4$&$(12092, 12116, 12140)$ &$\phi$ & $4$/$-4$
\end{tabular}
\end{ruledtabular}
\end{table*}

\begin{table*}
\caption{The group theory factor $F$, defined in Eq.~\eqref{eq:potential}, for the interaction of bottom-antibottom/bottom-bottom hadron pairs with only the light vector-meson exchanges. Here one of the bottom hadrons is an $s_\ell=3/2$ bottom meson. See the caption of Table~\ref{tab:potentialsc}. } \label{tab:potentialsb1}
\centering
\begin{ruledtabular}
\begin{tabular}{l|ccccc}
System &  $I$&$S$ & Thresholds [MeV] & Exchanged particles & $F$ \\
\hline
$B^{(*)}\bar B_{1,2}$/$B^{(*)} B_{1,2}$& 0&0/0&$
(11005, 11051, 11018, 11063)$ &$\rho,\omega$ & $\frac32,\frac12$/$\frac32,-\frac12$\\
& 1&0/0& & & $-\frac12,\frac12$/ $-\frac12,-\frac12$\\
$B^{(*)}\bar B_{s1,s2}$/$B^{(*)} B_{s1,s2}$& $\frac12$&$1/-1$&$(11093, 11141, 11105, 11154)$&$-$ & $0$/$0$\\
$B_s^{(*)}\bar B_{1,2}$/$B_s^{(*)} B_{1,2}$& $\frac12$&$-1/1$&$(11108, 11153, 11119, 11165)$&$-$ & $0$/$0$\\
$B^{(*)}_s\bar B_{s1,s2} $/$B^{(*)}_s B_{s1,s2} $&0& 0/$-2$&$(11196, 11207, 11244, 11255)$&$\phi$ & $1$/$-1$\\
\hline
$B_{1,2}\bar B_{1,2}$/$B_{1,2} B_{1,2}$&0&0/0&$(11452, 11464, 11477)$ &$\rho,\omega$ & $\frac32,\frac12$/$\frac32,-\frac12$\\
& 1&& & & $-\frac12,\frac12$/$-\frac12,-\frac12$\\
$B_{s1,s2}\bar B_{1,2}$/$B_{s1,s2} B_{1,2}$& $\frac12$&1/1&$(11555, 11566, 11567, 11578)$& & $0$/0\\
$B_{s1,s2}\bar B_{s1,s2} $/$B_{s1,s2} B_{s1,s2} $&0& $0/-2$&$(11657, 11669, 11680)$&$\phi$ & $1$/1\\
\hline
$\Lambda_b\bar B_{1,2}$/$\Lambda_b B_{1,2}$& $\frac12$&0/0 &$(11346, 11358)$&$\omega$ & $-1$/1\\
$\Lambda_b\bar B_{s1,s2}$/$\Lambda_b B_{s1,s2}$&0& $-1/1$ &$(11448, 11459)$&$-$ & $0$/0\\
$\Xi_b\bar B_{1,2}$/$\Xi_b B_{1,2}$& 1&$-1/-1$ &$(11520, 11533)$ &$\rho,\omega$ & $-\frac12,-\frac12$/$-\frac12,\frac12$\\
& 0& & & & $\frac32,-\frac12$/$\frac32,\frac12$\\
$\Xi_b\bar B_{s1,s2}$/$\Xi_b B_{s1,s2}$& $\frac12$&$-2/0$& $(11623, 11634)$&$\phi$ & $-1$/1\\
\hline
$\Sigma_b^{(*)}\bar B_{1,2}$/$\Sigma_b^{(*)} B_{1,2}$& $\frac32$&0/0&$(11539, 11551, 11559, 11571)$ &$\rho,\omega$ & $-1,-1$/$-1,1$\\
& $\frac12$&& & & $2,-1$/2,1\\
$\Sigma_b^{(*)}\bar B_{s1,s2}$/$\Sigma_b^{(*)} B_{s1,s2}$&1& $1/-1$&$(11642, 11653, 11661, 11672)$ &$-$ & $0$/0\\
$\Xi_b^{'(*)}\bar B_{1,2}$/$\Xi_b^{'(*)} B_{1,2}$&1& $-1/-1$&$(11661, 11673, 11680, 11692)$ &$\rho,\omega$ & $-\frac12,-\frac12$/$-\frac12,\frac12$\\
& 0&& & & $\frac32,-\frac12$/$\frac32,\frac12$\\
$\Xi_b^{'(*)}\bar B_{s1,s2}$/$\Xi_b^{'(*)}B_{s1,s2}$& $\frac12$&$-2/0$& $(11764, 11775, 11783, 11794)$&$\phi$ & $-1$/1\\
$\Omega_b^{(*)}\bar B_{1,2}$/$\Omega_b^{(*)} B_{1,2}$& $\frac12$&$-2/-2$&$(11772, 11784, 11796, 11808)$ &$-$ & $0$/0\\
$\Omega_b^{(*)}\bar B_{s1,s2}$/$\Omega_b^{(*)} B_{s1,s2}$&0& $-3/-1$&$(11875, 11886, 11899, 11910)$ &$\phi$ & $-2$/2\\
\end{tabular}
\end{ruledtabular}
\end{table*}
\medskip

\bibliography{ref-update}

\end{document}